\documentclass[%
 reprint,
superscriptaddress,
 amsmath,amssymb,
 aps,
prx,
]{revtex4-2}
\usepackage{color,soul}
\usepackage{graphicx}
\usepackage{dcolumn}
\usepackage{bm}
\usepackage{siunitx}
\usepackage{hyperref}
\usepackage{lipsum}
\usepackage{mhchem}
\usepackage{xcolor}
\usepackage{soul}
\usepackage{changes}
\usepackage[utf8]{inputenc}
\usepackage[T1]{fontenc}

\newcommand{\EPFL}{Swiss Federal Institute of Technology Lausanne (EPFL), CH-1015 Lausanne, Switzerland}

\newcommand{\CTH}{Department of Microtechnology and Nanoscience (MC2),
	Chalmers University of Technology, SE-412 96 G\"oteborg, Sweden}

\newcommand{\fref}[1]{Fig. \ref{#1}}
\newcommand{\subfref}[2]{Fig. \ref{#1}(#2)}
\newcommand{\eref}[1]{Eq. (\ref{#1})}
\newcommand{\appref}[1]{Appendix \ref{#1}}
\newcommand{\vecr}{\vec{r}}
\newcommand{\vecrp}{\vec{r}\,'}

\newcommand{\Kcoul}{\frac{1}{4\pi\epsilon_0}}

\begin{document}

\preprint{APS/123-QED}




\title{Noncontact friction in ultracoherent nanomechanical resonators near dielectric materials}


\author{Amirali Arabmoheghi}
\email{amirali.arabmoheghi@epfl.ch}
\affiliation{\EPFL}

\author{Alessio Zicoschi}
\affiliation{\EPFL}

\author{Guillermo Arregui}
\affiliation{\EPFL}

\author{Mohammad J. Bereyhi}
\affiliation{\EPFL}

\author{Yi Xia}
\affiliation{\EPFL}

\author{Nils J. Engelsen}
\affiliation{\EPFL}
\affiliation{\CTH}

\author{Tobias J. Kippenberg}
\email{tobias.kippenberg@epfl.ch}
\affiliation{\EPFL}

\date{\today}

\begin{abstract}
\end{abstract}

\maketitle
\begin{figure*}[t!]
\includegraphics{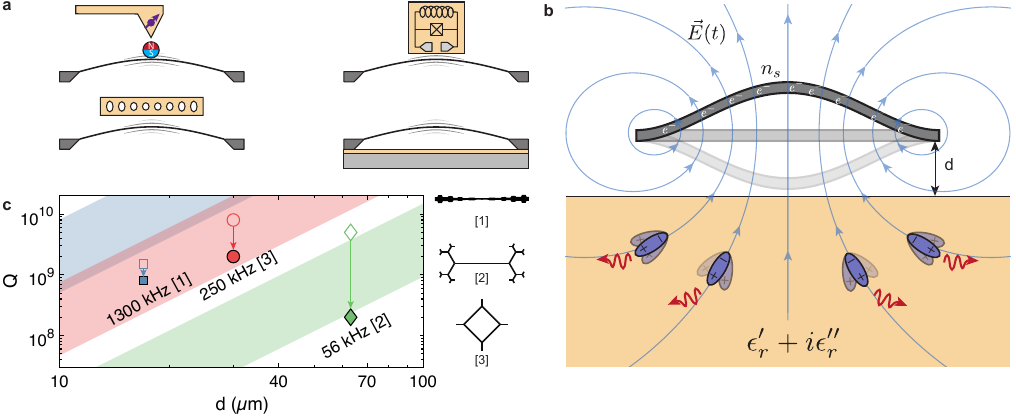}
\caption{\textbf{Noncontact friction in nanomechanical resonators near dielectrics a} Schematics of platforms for coupling nanomechanical resonators to spins in solids, superconducting circuits and nanophotonics cavities. In these platforms there are dielectric materials close to the resonator. Even in the absence of any external objects, the dielectrics on the substrate can also introduce loss. \textbf{b} Concept of dielectric-induced mechanical loss. The motion of the charged nanomechanical resonator induces a time-varying electric field within the dielectrics in its proximity, which cuases dissipation due to dielectric loss. \textbf{c} Dielectric loss within the substrate limiting the quality factor of ultracoherent strings from Ref. \cite{ghadimi_elastic_2018, bereyhi_hierarchical_2022, bereyhi_perimeter_2022}. The colors blue, red and green correspond to strained-engineered, hierarchical and polygon designs, shown in the right hand side. The ribbons correspond to the estimated NCF-limited quality factor for the frequency and geometry of each design, suspended above a silicon substrate with a \SI{4}{\nano\meter} native \ce{SiO2} layer. The surface charge density is in the \SI{1.45e11}{} to \SI{5.8e11}{e/\centi\meter^2} range. The filled and empty markers correspond to the measured and simulated quality factors adapted from the references.}
\label{fig:concept}
\end{figure*}

\textbf{Micro- and nanomechanical resonators are emerging as promising platforms for quantum technologies, precision sensors and fundamental science experiments.
To utilize these devices for force sensing or quantum optomechanics, they must be brought in close proximity with other systems for functionalization or efficient readout.
Improved understanding of the loss mechanisms in nanomechanical resonators, specifically the advent of dissipation dilution, has led to the development of resonators with unprecedented coherence properties.
The mechanical quality factors of this new class of ultracoherent micro- and nanomechanical oscillators can now exceed 1 billion at room temperature, setting their force sensitivities below \SI{1}{\atto\newton/\sqrt{\hertz}}, surpassing those of the state-of-the-art atomic force microscopes (AFMs).
Given this new regime of sensitivity, an intriguing question is whether the proximity of other materials hinders mechanical coherence.
Here we show: it does. We report a novel dissipation mechanism that occurs in ultracoherent nanomechanical oscillators caused by the presence of nearby dielectrics. By studying the parameter scaling of the effect, we show that the mechanism is more severe for low-frequency mechanical modes and that it is due to dielectric loss within the materials caused by the motion of a resonator which carries static charges. Our observations are consistent with the noncontact friction (NCF) observed in AFMs. Our findings provide insights into limitations on the integration of ultracoherent nanomechanical resonators and highlight the adverse effects of charged defects in these systems.}

\section{Introduction}
Micro- and nanomechanical oscillators have long been used in science and technology alike, as they can couple to a wide range of other degrees of freedom. However, their miniaturization renders them more susceptible to dissipative mechanisms. Reaching quality factors beyond $10^9$, comparable to those in bulk systems \cite{braginsky_systems_1987}, was not achieved in micro- and nanomechanical resonators until recently \cite{maccabe_nano-acoustic_2020, beccari_strained_2022, bereyhi_perimeter_2022, cupertino_centimeter-scale_2024, shin_spiderweb_2022,rossi_measurement-based_2018}.
A large family of ultrahigh-Q resonators is based on flexural modes of tensioned and high-aspect-ratio structures (i.e. membranes and strings) \cite{engelsen_ultrahigh-quality-factor_2024}.
In these resonators dissipation dilution, a phenomenon first introduced in the gravitational wave detector community \cite{huang_dissipation_1998}, enhances the quality factor of the modes by many orders of magnitude beyond the material’s intrinsic quality factor \cite{fedorov_generalized_2019}. String-like resonators with soft-clamped modes \cite{ghadimi_elastic_2018, bereyhi_hierarchical_2022, bereyhi_perimeter_2022, shin_spiderweb_2022, cupertino_centimeter-scale_2024} are particularly promising force sensors as they exhibit both low mass ($m\sim$ \SI{}{\pico\gram}) and low damping rate ($\Gamma_m<$\SI{1}{\milli\hertz}). The thermal-limited force sensitivity of mechanical oscillators at temperature $T$ is commonly defined as the single-sided spectrum of the thermal force, i.e. $\sqrt{S_F^\mathrm{th}} = \sqrt{4k_BTm\Gamma_m}$, where $k_B$ is the Boltzmann's constant. For these ultracoherent devices, it reaches values below \SI{1}{\atto\newton/\sqrt\hertz} at room temperature \cite{bereyhi_perimeter_2022}, surpassing the sensitivity level of the state-of-the-art AFM cantilevers \cite{heritier_nanoladder_2018}.

Virtually all applications that require low mechanical dissipation rely on short-range interactions and require placing other objects at sub-micron distances from the nanomechanical resonator. For example, measurements of Casimir forces \cite{zou_casimir_2013}, tests of gravity at short distances \cite{arkanihamed_hierarchy_1998} and precision magnetic resonance force microscopy (MRFM) experiments \cite{halg_membrane-based_2021, gisler_enhancing_2024} all occur at short distances. Moreover, as illustrated in \subfref{fig:concept}{a}, several platforms for coupling nanomechanical resonators to other degrees of freedom, such as spin qubits \cite{fung_toward_2024, rabl_quantum_2010}, superconducting circuits \cite{seis_ground_2022} and optical microcavities \cite{guo_active-feedback_2023, guo_integrated_2022, guo_feedback_2019, anetsberger_near-field_2009, gavartin_hybrid_2012} also rely on near-field interactions. The latter is of particular interest as the optomechanical coupling, which enables readout of the motion of the resonator with quantum-limited precision \cite{aspelmeyer_cavity_2014}, scales exponentially with the distance between optical and mechanical resonators. 
So far, the devices used in these applications \cite{halg_membrane-based_2021, gisler_enhancing_2024,fung_toward_2024, rabl_quantum_2010,guo_active-feedback_2023, guo_integrated_2022, guo_feedback_2019, anetsberger_near-field_2009, gavartin_hybrid_2012} do not exhibit the force sensitivities, $S_F^\mathrm{th}$, as low as those of the \emph{stand-alone} ultrahigh-$Q$ devices discussed previously, which are suspended tens of micrometers away from the closest object. This naturally raises the question whether ultracoherent mechanical devices, with force sensitivity below that of AFMs, can retain their coherence properties near other materials.

In this work, we show that nearby dielectrics do perturb nanomechanical resonators. We report on an unexpected loss mechanism occurring when ultracoherent string resonators are integrated in the near-field of a photonic crystal (PhC) cavity.
We observe that the $Q$ of the soft-clamped mode of a binary tree-shaped resonator \cite{bereyhi_hierarchical_2022} is reduced for smaller cavity-resonator separation. Moreover, for millimeter-long strings, even the presence of the substrate supporting the structures can cause additional loss in the low-frequency modes. By identifying that the added loss is inversely proportional to the mechanical mode's frequency, we realize that it is due to an electrical interaction between the resonator and the dielectrics in its proximity (i.e. the optical cavity or the substrate). This phenomenon, also known as noncontact friction (NCF), has only been observed in atomic force microscopes (AFMs) cantilevers \cite{stipe_noncontact_2001, yazdanian_dielectric_2008, poggio_force-detected_2010} at nanometer-scale tip-sample distances. In nanomechanical systems, outside the context of AFMs, these effects have been largely overlooked due to larger inter-body distances and the fact that most resonators are not intentionally charged. However, the presence of trapped charges on micro-fabricated devices, in particular with amorphous materials, has been commonly reported \cite{baik_charge_2021, schmidt_influence_2007, krick_nature_1988, sharma_study_2013, sonnefeld_determination_1996}. We have developed a theoretical and numerical framework for studying NCF in nanomechanical systems in the presence of static charges and we show that they explain our observations very well. In isolated devices with ultracoherent soft-clamped modes, despite the inverse scaling of the highest Q with the mode frequency, i.e. $Q\propto \Omega^{-2}$, experimental realizations fail to follow this trend below certain frequencies \cite{ghadimi_elastic_2018, bereyhi_hierarchical_2022, bereyhi_perimeter_2022} and this reduction in Q for low-frequency modes has not been understood. We show that NCF due to the presence of the substrate, even tens of microns away, could explain the discrepancy (see \subfref{fig:concept}{c}).

\section{Dielectric-induced mechanical loss}\label{sec:Theory}
\begin{figure*}[t!]
    \centering
    \includegraphics{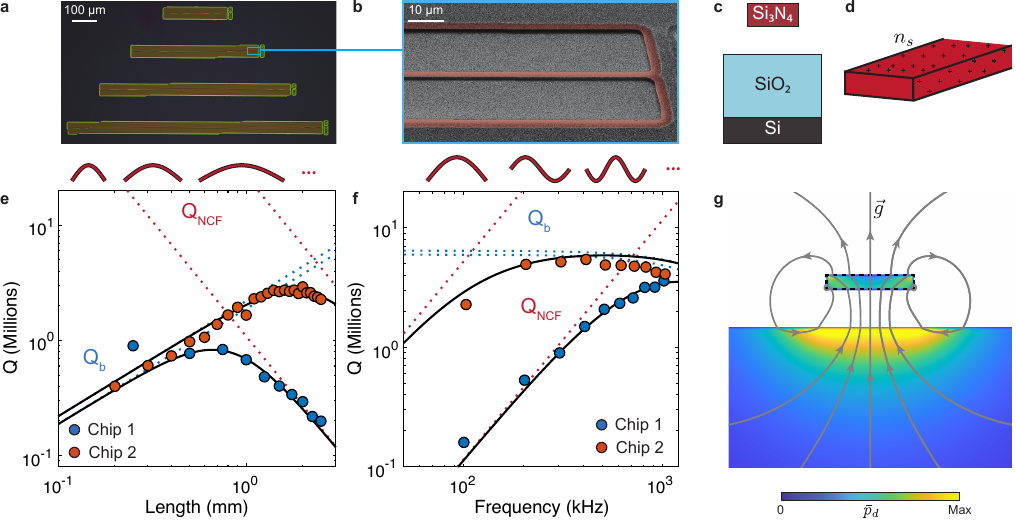}
    \caption{\textbf{Noncontact friction in strings suspended above a dielectric surface. a} Optical microscope image of a few uniform doubly clamped beams suspended above the \ce{SiO2} layer. \textbf{b} False colored scanning electron microscope (SEM) micrograph of a suspended nano-beam taken close to its clamping point. Red: suspended \ce{Si3N4}. \textbf{c} Full stack of the structures after the release. \textbf{d} Rectangular cross-section of the nano-beams and different charge distributions on the horizontal and vertical surfaces. \textbf{e} Measured $Q$ for the fundamental modes of nano-beams with different lengths. Circles: experimental data from chips 1 and 2. Black solid lines: fits from the $Q_{\mathrm{tot},n}$ model. Blue and red dotted lines correspond to the bending and NCF components of the fits respectively. \textbf{f} Measured $Q$ for the higher order modes of the \SI{2.5}{\milli\meter} long beams. Markers, lines and colors denote the same as in \textbf{e}. \textbf{g} FEM simulation of NCF. The dashed lines on the edges of the string indicate the charged surfaces. The string is \SI{1.5}{\micro\meter} wide and \SI{227}{\nano\meter} thick. The gap is \SI{680}{\nano\meter}. Grey lines: the streamlines of the $\vec g$ field. The colormap shows the dissipated power density.}
    \label{fig:NCF_in_ToyModel}
\end{figure*}

Noncontact friction has been modeled in AFM systems as the interaction of a moving point charge with a lossy dielectric \cite{yazdanian_dielectric_2008}. This model can be readily generalized to nanomechanical resonators with distributed electric charges. As shown in \subfref{fig:concept}{b}, a nanomechanical resonator (e.g., a doubly clamped string), with a volume charge density $\rho_e(\vecr)$, is placed near a rigid dielectric body with the complex frequency-dependent permittivity $\epsilon_r(\vecr, \omega) = \epsilon_r'(\vecr, \omega)+i\epsilon_r''(\vecr, \omega)$. $\epsilon''_r(\vecr, \omega)$ characterizes the material's losses (which generally include both dielectric and conductive components). In the equilibrium position, the electric field in the space is $\vec E_0(\vecr)$. Consider a vibrational eigenmode of the resonator with frequency $\Omega_m$ and mode shape $\vec U_m(\vecr)$. Displacements in this mode, $u(t) = x(t)e^{-i\Omega_mt} + c.c.$ with slowly varying amplitude $x(t)$, modify the electric field as
\begin{equation}\label{eq:field_modif}
    \vec E(\vecr,t) = \vec E_0(\vecr) + \vec g(
    \vecr, \Omega_m;\vec U_m(\vecr))u(t),
\end{equation}
where $\vec g(\vecr, \Omega_m;\vec U_m(\vecr)) = \frac{\delta\vec E_0(\vecr)}{\delta u}$ is the electric field variations under the displacements of the mode $\vec U_m$ (see \appref{app:detailed_theory} for details). Due to the imaginary part of $\epsilon_r$, the motion of the resonator results in dissipation within the dielectric with the period-averaged dissipated power density $\bar p_d(\vecr) = \frac{\Omega_m}{2}\epsilon_0\epsilon''_r(\vecr, \Omega_m)|\vec g|^2 |x|^2$. Integrating over all dielectrics gives the period-averaged total power dissipated by the oscillator. Referring the loss to a viscous damping force of the form $-\gamma_\mathrm{NCF}\dot u(t)$, the dissipated power would be $\bar P_d = \frac{\Omega_m^2}{2}\gamma_\mathrm{NCF}|x|^2$. Comparing these expressions yields the NCF damping coefficient
\begin{equation}\label{eq:gamma_ncf}
    \gamma_\mathrm{NCF} = \int d^3r \,\epsilon_0|\vec g(\vecr, \Omega_m;\vec U_m(\vecr))|^2\frac{\epsilon''_r(\vecr, \Omega_m)}{\Omega_m}.
\end{equation}
For insulators with frequency-independent losses, the inverse-frequency scaling ($\gamma_\mathrm{NCF}\propto 1/\Omega_m$) provides a clear experimental signature of NCF in nanomechanics.

Using this formulation and a geometry similar to \subfref{fig:concept}{b}, for ultracoherent silicon nitride strings suspended above silicon substrates, we estimate the quality factor imposed by the dielectric loss within the thin native oxide layer on the substrate (\subfref{fig:concept}{a}) (see \appref{app:toy_model}). The results are shown as a function of the string-substrate distance in \subfref{fig:concept}{c}, assuming the devices to have a range of typical surface charge densities \cite{krick_nature_1988}. To compare to experimental results, we choose three devices with elastic strain engineering \cite{ghadimi_elastic_2018}, hierarchical \cite{bereyhi_hierarchical_2022} and polygon \cite{bereyhi_perimeter_2022} designs that exhibited quality factors that were significantly below the simulation values. We have measured the string-substrate distances on these devices and compute the NCF-limited $Q$ for their geometry and the corresponding mode frequencies. As shown in \subfref{fig:concept}{c}, despite the large variations, the NCF damping within the native oxide layer could explain the reduction of $Q$ in these samples.

\section{Noncontact friction in strings suspended above a dielectric surface}\label{sec:iii}
For the simple system shown in \subfref{fig:concept}{b}, we can find an analytical expression for the NCF damping coefficient. The string is infinitesimally thin and carries linear charge density $\rho_{e,l}$. For this system, we can find the electric field variation using the elementary method of image charges (see \appref{app:toy_model}). The NCF-limited quality factor for mode order $n$ with frequency $\Omega_n$ is given by
\begin{equation}\label{eq:Q_ncf_n}
    Q_{\mathrm{NCF},n} = \frac{8\pi\epsilon_0\rho_{m,l}d^2}{\rho_{e,l}^2}\frac{\Omega_n^2}{\mathrm{Im}[\xi(\Omega_n)]},
\end{equation}
where $\xi(\omega) = \frac{\epsilon_r(\omega)-1}{\epsilon_r(\omega)+1}$ and $\rho_{m,l}$ is the linear mass density. To experimentally study this model, we fabricated uniform strings, suspended above a dielectric substrate (overview shown in \subfref{fig:NCF_in_ToyModel}{a,b}). As shown in \subfref{fig:NCF_in_ToyModel}{c}, the substrate consists of a thick \ce{SiO2} layer and the strings are made of high-stress ($\sim$\SI{1}{\giga\pascal}) \ce{Si3N4} with a rectangular cross-section (see \subfref{fig:NCF_in_ToyModel}{d}). Fabrication details are provided in \appref{app:fabrication_details}. The devices have lengths ranging from 200 to 2500 \SI{}{\micro\meter} and are suspended \SI{500}{\nano\meter} above the substrate. We measure the quality factor of their out-of-plane (OP) modes, under vacuum ($P<10^{-8}$ mbar), through ringdown measurements using an optical interferometer \cite{ghadimi_elastic_2018, bereyhi_hierarchical_2022, bereyhi_perimeter_2022}. For these modes, when no dielectrics are nearby, the quality factor is limited by bending loss. For a string of thickness $h$ and length $L$, it is given by the dissipation-dilution theory \cite{fedorov_generalized_2019}
\begin{equation}\label{eq:Q_int_n}
Q_{b,n} = \frac{Q_\mathrm{int}}{2\lambda + n^2\pi^2\lambda^2},
\end{equation}
where $Q_\mathrm{int}$ is the intrinsic mechanical $Q$ of the material, and $\lambda = \sqrt{E/12\sigma},h/L$, with $E$ and $\sigma$ being the Young's modulus and stress, respectively.

We measure the $Q$ of the fundamental modes of strings with varying lengths on two different chips from separate fabrication runs. As shown in \subfref{fig:NCF_in_ToyModel}{e}, the quality factors significantly deviate below the bending loss model of \eref{eq:Q_int_n}. We also measure the $Q$ of the first 10 OP modes of the \SI{2500}{\micro\meter} long beams from both chips. According to \eref{eq:Q_int_n}, since $\lambda\ll 1$, we expect a weak descending dependence on the mode frequency. However, the data shown in \subfref{fig:NCF_in_ToyModel}{f} indicate that modes with lower frequencies experience more damping.

The observations can be explained well by a model that accounts for both bending loss and NCF: $Q_{\mathrm{tot},n}^{-1} = Q_{b,n}^{-1} + Q_{\mathrm{NCF},n}^{-1}$. In \subfref{fig:NCF_in_ToyModel}{e} and (f), we show the results of fitting with $Q_\mathrm{int}$ and the product $\rho_{e,l}^2\mathrm{Im}[\xi]$ as free parameters. Note that despite the finite rectangular cross-section of the strings, we can define an effective linear charge density, $\rho^\mathrm{eff}_{e,l}$ (see \appref{app:FEM_NCF}). By using the literature values for \ce{SiO2} permittivity and loss tangent, we extract this effective linear charge density. The analysis of the fundamental modes yields the $\rho^\mathrm{eff}_{e,l}$ of \SI{180}{e/\micro\meter} and \SI{40}{e/\micro\meter} for devices from chip 1 and 2, respectively. From the analysis of the higher order modes we obtain \SI{210}{e/\micro\meter} and \SI{42}{e/\micro\meter}, in agreement with the former analysis within 15\%. In \appref{app:toy_model} we show that the correction needed for taking the effect of the conductive silicon substrate, for the thicknesses of the oxide layer (>\SI{1}{\micro\meter}) is less than 10\%.
Interestingly, while the bending contributions are consistent across chips, the NCF contributions differ drastically. This observation suggests that the charge density is highly fabrication-run-dependent. Nevertheless, the good agreement between the fits and the data is strong evidence that the additional loss channel observed in this system is indeed NCF. To the best of our knowledge, no other relevant damping mechanism can account for the observed $\Omega_m^2$ scaling in $Q$. (see \appref{app:rule_out_others} for a discussion).

To go beyond the infinitesimally thin string approximation, we developed an FEM model to compute the field variation $\vec g$ and NCF coefficient for arbitrary geometries (see \appref{app:FEM_NCF}). \subfref{fig:NCF_in_ToyModel}{g} shows the result of this simulation for out-of-plane motion of a rectangular string. Since the mode shape varies slowly compared to the gap size, the simulation is done in the 2D cross-section of the geometry and the result is analytically obtained for the full geometry (see \appref{app:3Dto2D}).
We have computed $\vec g$ and the dissipated power density, $\bar p_d$, within both the substrate and the string itself.
For simplicity, we have assumed a uniform charge distribution on all beam surfaces.
Using our simulation, we can obtain the surface charge density from the effective linear ones mentioned earlier. These values amount to \SI{8.3e9}{} and \SI{1.2e9}{e/\centi\meter^2} for chips 1 and 2, respectively. Here, unlike the simulation shown in \subfref{fig:NCF_in_ToyModel}{g}, we have included the conductive silicon substrate in the FEM simulation as well. These charge densities are smaller than typically reported values in the range of $10^{11}-10^{13}$ \SI{}{e/\centi\meter^2} \cite{sharma_study_2013, sonnefeld_determination_1996, krick_nature_1988}. However, reported values vary widely and depend on the specific chemical treatment used in each case \cite{sonnefeld_determination_1996, lin_surface_2021}.
In an attempt to understand the origin of the charges, we have gathered more data showing a higher density on the vertical sidewalls (see \appref{app:width_dependence}). While not conclusive, the data hints towards the effect of plasma etching on the charge densities.

\section{Observation of noncontact friction in a nano-optomechanical system}
\begin{figure*}[t!]
    \centering
    \includegraphics{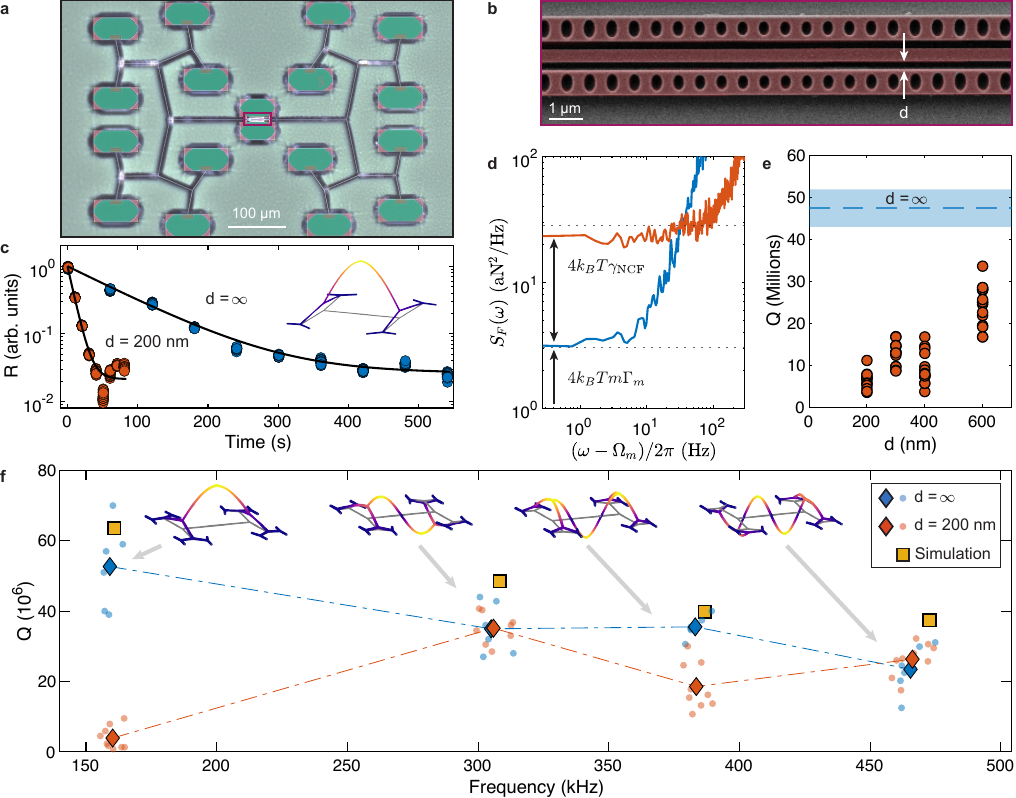}
    \caption{\textbf{Noncontact friction in a nano-optomechanical system. a} Optical microscope image of the nano-optomechanical system. The suspended binary-tree resonators can be seen as the thin white segments. The green boxes are clamping points. The PhC cavity is located at the center of the device. \textbf{b} False-colored SEM micrograph of the section marked by the red rectangle in \textbf{a}, taken prior to the release, showing the PhC cavities and the nano-beam in between them. \textbf{c} Gated ringdown measurements of the mode with the shown mode shape in an only-beam device (blue) and integrated device with gap of \SI{200}{\nano\meter} (orange). \textbf{d} Calibrated force noise PSD for the same modes measured in \textbf{c}. \textbf{e} Compilation of the measured $Q$s for integrated with varying gaps (orange circles) as well as the average of the only-beam devices (the dashed blue line). The blue shade corresponds to the standard deviation of the only-beam $Q$s. Each circle corresponds to the measurement from one device. \textbf{f} Measured $Q$ for the first 4 OP modes with the shown mode shapes, for only-beam (blue markers) and integrated (orange markers) devices with the gap of \SI{200}{\nano\meter}. Circles show single devices and diamonds the average value. The dash-dotted lines are guides to the eye. Yellow squares: FEM simulation values.}
    \label{fig:NCF_in_OM}
\end{figure*}
\begin{figure*}
    \centering
    \includegraphics{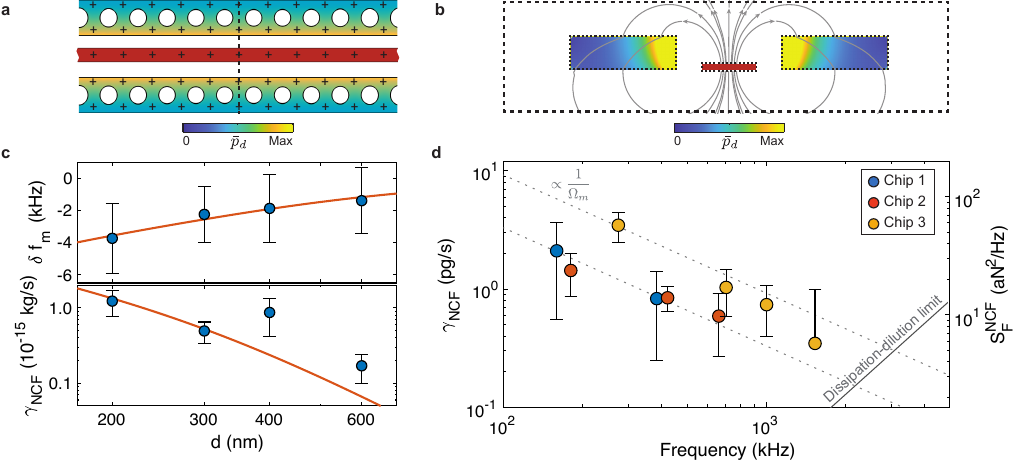}
    \caption{\textbf{Modeling noncontact friction in a nano-optomechanical system. a} Simulation of the dissipated power in the PhC cavities due to NCF. A zoom out of the central part of \subfref{fig:NCF_in_OM}{b}. The plus marks indicate the static charge distribution. The colormap shows the dissipated power. \textbf{b} Cross-section of \textbf{a} alongside the field lines corresponding to the electric field variation $\vec g$. The dashed outlines indicated the charged surfaces. \textbf{c} top: Frequency shift of the fundamental modes of integrated devices with varying gaps referenced to the average only-beam value. Blue circles: average value for devices with same gap. Error bar corresponds to the standard deviation of the ensemble. Solid orange line: FEM simulations with the charge density and loss tangent obtained from the fits. Shaded area: 95\% confidence interval of these values. Bottom: NCF-limited damping coefficients for the same devices in the top panel. \textbf{d} NCF-limited damping coefficient for odd order modes of an ensemble of 77 integrated devices on three different chips. Dotted gray lines are guides to eye for $1/\Omega_m$ scaling. The right axis corresponds to the same values calibrated in the force noise PSD units. The gray solid line is the lower limit of $S_F^\mathrm{th}$ for soft-clamped modes.}
    \label{fig:NCF_OM_analysis}
\end{figure*}
We have implemented on-chip integration of a \ce{Si3N4} photonic crystal (PhC) microcavity with a binary-tree string resonator \cite{bereyhi_hierarchical_2022} (see \appref{app:fabrication_details} for the fabrication process). A representative device is shown in \subfref{fig:NCF_in_OM}{a} and (b). The string is suspended in the gap between two identical PhC cavities. The coupling of the string's motion to the optical mode allows for efficient displacement readout. The design and performance of the optomechanical device are not the main focus of this work, and we focus on the loss introduced by the presence of the PhC cavities. On each chip, we fabricate devices with the string-PhC gap size, $d$, ranging from 150 to \SI{600}{\nano\meter}, as well as reference devices without any PhC cavities (i.e. $d=\infty$). From here on, we refer to the devices featuring both the string and the PhCs as the ``integrated'' devices and the ones without the PhCs as the ``reference'' devices. As in the previous section, we measured the $Q$ for the OP modes of the resonators using a free-space interferometer. To avoid optomechanical effects in the near-field system, no light was coupled into the PhC cavities.

\subfref{fig:NCF_in_OM}{c} shows a ringdown measurement for the fundamental modes of two devices fabricated on the same chip with identical binary-tree string designs: an integrated device and a reference device. According to the simulations, this mode is expected to occur at \SI{160}{\kilo\hertz}, with $Q=5\times10^7$ and effective mass of $m_\mathrm{eff} = \SI{7.6}{\pico\gram}$. While good agreement is observed for the reference device, the integrated device with $d=\SI{200}{\nano\meter}$ has about ten times lower $Q$, $4.7\times 10^6$.
As shown in \subfref{fig:NCF_in_OM}{d}, a calibrated measurement of $S_F^\mathrm{th}$ \cite{bereyhi_perimeter_2022} also confirms an increase in $S_F^\mathrm{th}$ from 3 to \SI{21}{\atto\newton^2/\hertz}, consistent with the ringdown results. We further characterize an ensemble of 77 reference and integrated devices with varying gaps, all on the same chip. The data shown in \subfref{fig:NCF_in_OM}{e} shows a systematic reduction of the $Q$ for shorter gaps. In binary-tree resonators, in addition to the fundamental mode, the higher order OP modes are also soft-clamped and possess high quality factors. We study reference and integrated devices with $d = \SI{200}{\nano\meter}$ to compare the first four OP modes. FEM simulations of the mode shapes and the outcome of the measurements are shown in \subfref{fig:NCF_in_OM}{f}. For the fundamental mode the contrast between the reference and integrated devices is reproduced. Although the third-order mode has a similar behavior, even-order modes that have a node at the center experience no significant $Q$ reduction. This observation confirms that the additional loss channel arises from a local interaction with the PhC cavity.

To model NCF in our system, we consider a uniform charge distribution on all the surfaces of the device. This includes the surfaces of both the string and the PhC cavities (see \subfref{fig:NCF_OM_analysis}{a} and (b)). A key observation supporting the idea that the string and PhCs carry charges of the same sign is the gap-dependent frequency shift seen in the devices of \subfref{fig:NCF_in_OM}{e}. In \subfref{fig:NCF_OM_analysis}{c}, we show this frequency shift relative to the average of the reference devices. The observed negative frequency shift at smaller gaps arises from Coulomb repulsion between like charges on the string and the PhC. We calculate the frequency shift as a function of the gap using an FEM model (see \appref{app:FEM_NCF}) and fit the model to the data, using the surface charge density as the only free parameter. The fit, shown in \subfref{fig:NCF_OM_analysis}{c}, yields $\rho_{e,s} = 8.8 \pm0.9\times 10^9$ \SI{}{e/\centi\meter^2}, in good agreement with the value obtained in the previous section. In \subfref{fig:NCF_OM_analysis}{a} and (b), we also show the dissipated power due to NCF, computed using our FEM model. For better data analysis and comparison with this model, for every mechanical mode we introduce the experimentally inferred NCF damping coefficient as
\begin{equation}\label{eq:gamma_NCF_exp}
    \tilde\gamma_\mathrm{NCF} = m_\mathrm{eff}(\Gamma_m^{d} - \langle\Gamma_m^{\infty}\rangle),
\end{equation}
where $m_\mathrm{eff}$ is the effective mass of the mode, given by the FEM simulation, $\Gamma_m^{d}$ is the damping rate of the mode for the integrated device with gap $d$ and $\langle\Gamma_m^{\infty}\rangle$ is the average value of the damping rates of the same mode in the reference devices of the same chip. For the data in \subfref{fig:NCF_in_OM}{e}, we plot $\tilde\gamma_\mathrm{NCF}$ versus gap in \subfref{fig:NCF_OM_analysis}{c}. Using the charge density extracted from the frequency shift analysis, we fit the FEM model to the data, with the dielectric loss tangent of \ce{Si3N4}, $\tan\delta_\mathrm{\ce{Si3N4}}$, as the only free parameter. The result is shown in \subfref{fig:NCF_OM_analysis}{c} and yields $\tan\delta_\mathrm{\ce{Si3N4}}= 9\pm 3 \times 10^{-5}$. To our knowledge, the measurement of this quantity for thin-film \ce{Si3N4} at sub-megahertz frequencies and room temperature has not been performed before. We emphasize that making this measurement using electrical circuits is not straightforward due to ohmic losses.

The inferred NCF damping $\tilde\gamma_\mathrm{NCF}$ allows us to extract the frequency dependence of this loss mechanism using the higher modes of the binary-tree resonators, despite differences in their effective masses. \subfref{fig:NCF_OM_analysis}{d} shows $\tilde\gamma_\mathrm{NCF}$ for the odd-order modes of three different mechanical designs across three chips. All devices have a gap of $d=\SI{200}{\nano\meter}$ and all mode shapes vary slowly near the PhCs, allowing direct comparison of modes from different chips. We observe a $1/\Omega_m$ trend for devices on each chip, consistent with theoretical expectations (see \appref{app:rule_out_others} for a discussion). Devices on chip 3 exhibit higher NCF than those on chips 1 and 2. Chips 1 and 2 come from the same fabrication run, whereas chip 3 was released in a separate run and exposed to hydrofluoric acid in the final stage (see \appref{app:fabrication_details}). We attribute this difference to variations in surface charge density, which can depend on exposure to different pH environments \cite{sonnefeld_determination_1996}. To contextualize the values in \subfref{fig:NCF_OM_analysis}{d}, we convert them to the PSD of the force noise as $S_F^\mathrm{NCF} = 4k_BT\tilde\gamma_\mathrm{NCF}$, representing a limit for force sensing with these devices. While increasing the mode frequency reduces the impact of NCF, it comes with a trade-off for the bending loss.
The ultimate limit of soft-clamping, or the clampless limit \cite{fedorov_generalized_2019}, occurs when the bending loss at the clamping points is fully suppressed. At this limit, the Q has the $1/\Omega_m^2$ scaling with the mode frequency.
Using an assumption for the effective mass (see \appref{app:dissipation_dilution_limit_to_S_F}) we obtain a lower bound for the $S_F^\mathrm{th}$ of soft-clamped modes, also shown in \subfref{fig:NCF_OM_analysis}{d}. Given the amount of charge on the devices, there is an optimal frequency at which the sensor has the best sensitivity.
\section{Conclusion}
We have shown that the presence of static charges on nanomechanical resonators gives rise to NCF when they are placed near other bodies, hindering their performance in tasks that require high quality factors.
Although we made simplifying assumptions about the charge distribution, our methods allow precise modeling of NCF in arbitrary geometries with arbitrary charge distributions, which can be characterized using standard techniques such as Kelvin probe force microscopy. In a broader context, our study highlights the challenges that static charges pose for microfabricated devices used in sensitive experiments.
Effects similar to NCF have been observed in other electric-field sensitive experiments such as trapped ions \cite{kumph_electric-field_2016} and Ryberg atoms \cite{ocola_control_2024}. With growing interest in scaling up hybrid quantum systems for computing, simulation, and sensing, microfabricated devices are inevitable. Understanding the limitations imposed by imperfections such as trapped charges is therefore essential. Conversely, our work points to opportunities for new functionalities: static charges can enable nanomechanical resonators to couple to electric fields and other quantum systems with electronic degrees of freedom, such as ions and Rydberg states of neutral atoms. Furthermore, the NCF mechanism itself can serve as a sensitive probe for dissipative processes in materials that imprint on the motion of mechanical oscillators. Our theoretical and numerical methods are readily applicable to these scenarios.

\bibliography{references.bib}
\begin{acknowledgments}
   The authors acknowledge Alexander Eichler for insightful discussions. This work was supported by funding from the Swiss National Science Foundation under grant agreement No.231403 (CoolMe) and No.216987 (CONSULT). All samples were fabricated in the Center of MicroNanoTechnology (CMi) at EPFL.
\end{acknowledgments}

\newpage
\mbox{}
\newpage

\begin{widetext}
\appendix
\section{Detailed theory of noncontact friction in deformed geometries}\label{app:detailed_theory}
The mathematical details of the computation of the NCF damping are discussed in this section. In particular, the properties of the electric field variation $\vec g$ and the methods for calculating it are discussed. We compute the dynamical effects on the moving bodies, meaning both frequency shifts and damping coefficients using two methods. The first is the same as the one discussed in the main text, i.e. computing the energy within the dielectrics. The second is to calculate the forces on the moving bodies.

\subsection{General properties of $\vec g$}
First of all, we limit the discussions in this paper to the situations where we can make the quasi-electrostatic approximation (i.e. size of the system $\ll \Omega/c$). Situations beyond this approximation are irrelevant to nanomechanics. The static electric field, for the undeformed geometry, obeys the equations,
\begin{subequations}\label{eq:Poissons}
    \begin{align}
    \nabla\cdot\left[\epsilon_r(\vecr)\vec E_0(\vecr)\right] &= \frac{\rho_e(\vecr)}{\epsilon_0},\\
    \nabla\times\vec E_0(\vecr) = 0,
\end{align}
\end{subequations}
\begin{figure}[t!]
\includegraphics{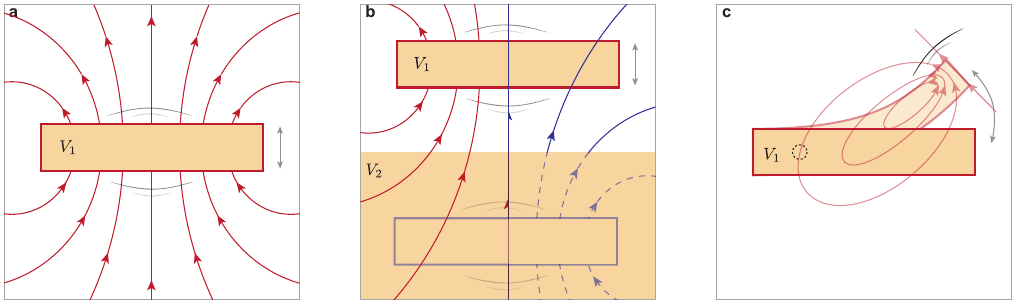}
\caption{\textbf{Illustration of the impact of relative motion on NCF a} Here an isolated dielectric with volume $V_1$ with charges on its surface (marked by red lines) has a vertical center of mass motion. In the laboratory frame, the field variation marked by the red streamlines is generated while at the reference frame of the dielectirc the field variation is zero as there is no motion. \textbf{b} In this case the charged dielectric $V_1$ moves above another dielectric with volume $V_2$. In the reference frame of $V_2$, the red field is generated, which is the one discussed in the main text and causes dissipation within $V_2$. However, in the reference frame of $V_1$, while the field variation due to the charges on the resonator (red charges) is  zero, the image charges in $V_2$ are in motion and generate the field variation shown by the blue lines within $V_1$. This field causes dissipation within the resonator. \textbf{c} Here the resonator is again isolated but instead of a center of mass motion, it is deformed (similar to a cantilever). In this scenario the field variation within the resonator also depends on the point of view. For example at an observation point towards the left side of the resonator (marked by the dashed circle) the motion of the charges on the other side of the resonator generate a field variation which generates loss within the resonator. This type of dissipation is present in any mechanical resonator that is charged, regardless of being isolated or not.}
\label{fig:app_relative_motion}
\end{figure}
as well as the usual boundary conditions for dielectric media, given by
\begin{subequations}
\begin{align}
    \left(\epsilon_{r}^{(2)}\vec E_{0}^{(2)} - \epsilon_{r}^{(1)}\vec E_{0}^{(1)}\right)\cdot\vec n &=\sigma_e, \label{eq:E_0_boundary_cond}\\
    \left(\vec E_{0}^{(2)} - \vec E_{0}^{(1)}\right)\times\vec n &= 0,
\end{align}
\end{subequations}
where $\sigma_e$ is the charge density at the boundary and $(1)$ and $(2)$ denote below and above it with respect to the normal vector $\vec n$. A deformation field $\vec u(\vecr, t)$, displaces both the charges and the dielectric boundaries. For small displacements, one can find the variation of $\rho_e(\vecr)$ and $\epsilon_r(\vecr)$ perturbatively by taking the transformation $\vecr\rightarrow\vecr+\vec u(\vecr,t)$. However, there is an important subtlety here. The mechanical-dielectric dynamical effects (both dispersive and dissipative) occur only when there is a relative motion between the charges and the dielectrics (see \fref{fig:app_relative_motion}). In other words, since the origin of the dynamical effects is the frequency dependent response of the dielectric, the field variation has to be calculated in the frame of reference of the dielectric. This makes the calculation of the field variation and eventually the dissipated power more complicated. To calculate the field variation at an observation point $\vecr$ in the reference frame of the dielectric at point $\vecrp$ we displace $\rho_e$ and $\epsilon_r$ using the transformation $\vecr\rightarrow\vecr+\vec\Delta(\vecr,\vecrp, t)$, where $\vec\Delta(\vecr,\vecrp, t) = \vec u(\vecr,t)-\vec u(\vecrp,t)$ is the relative displacement field. 

In the main text, for simplicity, we have neglected the relative motion effects and defined $\vec g$ as the generalized derivative of the static electric field, $\vec E_0$, with respect to the motion of a particular mode
\begin{equation}
    \vec g(\vecr, \Omega_m;\vec U_{m}) = \frac{\delta\vec E_0}{\delta u},
\end{equation}
where $u$ is the generic amplitude of the motion for mode $\vec U_m$. Here we use a perturbative approach and derive the equations governing $\vec g$.

We decompose the displacement field using the vibrational eigenmodes of the geometry. These modes are the solutions of the relevant elasticity equations and are characterized by the frequencies $\Omega_{m,\alpha}$ and the mode shapes $\vec U_{m,\alpha}(\vecr)$. The mode shapes are normalized to their maximum, i.e., $\max_{\vecr}|\vec U_{m,\alpha}(\vecr)| = 1$. The displacement field is given by
\begin{equation}\label{eq:u_rt_expansion}
    \vec u(\vecr, t) = \sum_\alpha x_\alpha\vec U_{m,\alpha}(\vecr)e^{-i\Omega_{m,\alpha}t} + \mathrm{c.c.},
\end{equation}
where $x_\alpha(t)$ is the amplitude of motion for the mode $\alpha$. The relative displacement field also reads
\begin{equation}
    \vec\Delta(\vecr,\vecrp, t) = \sum_\alpha x_\alpha\left[\vec U_{m,\alpha}(\vecr)-\vec U_{m,\alpha}(\vecrp)\right]e^{-i\Omega_{m,\alpha}t} + \mathrm{c.c.},
\end{equation}
Now we expand the perturbation in the charge density, permittivity and the electric field to the first order in $x_\alpha$
\begin{align}
    \rho_e(\vecr) &\rightarrow \rho_e(\vecr) + \sum_\alpha \frac{\delta \rho_e}{\delta x_\alpha} x_\alpha e^{-i\Omega_{m,\alpha}t} + \mathrm{c.c.},\\
    \epsilon_r(\vecr) &\rightarrow \epsilon_r(\vecr)+ \sum_\alpha \frac{\delta \epsilon_r}{\delta x_\alpha} x_\alpha e^{-i\Omega_{m,\alpha}t} + \mathrm{c.c.},\\
    \vec E_0(\vecr) &\rightarrow \vec E_0(\vecr) + \sum_\alpha\vec g(\vecr,\vecrp;\vec U_{m,\alpha})x_\alpha e^{-i\Omega_{m,\alpha }t} + \mathrm{c.c.},\label{eq:delta_E}
\end{align}
where
\begin{align}
    \frac{\delta \rho_e}{\delta x_\alpha } &= \nabla\rho_e(\vecr)\cdot\left[\vec U_{m,\alpha }(\vecr)-\vec U_{m,\alpha }(\vecrp)\right],\label{eq:rho_var_def}\\
    \frac{\delta \epsilon_r}{\delta x_\alpha } &= \nabla\epsilon_r(\vecr)\cdot\left[\vec U_{m,\alpha }(\vecr)-\vec U_{m,\alpha }(\vecrp)\right],\label{eq:eps_var_def}
\end{align}
and $\vec g(\vecr,\vecrp;\vec U_{m,\alpha })$ is the field variation under displacements by mode $\vec U_{m,\alpha }$ at coordinate $\vecr$, observed in the reference frame moving at coordinate $\vecrp$. We have not written the dependency on the frequency for the sake of notation simplicity. Plugging these relations in Equations \ref{eq:Poissons}
\begin{subequations}
    \begin{align}
        \sum_\alpha \left\{\nabla\cdot\left[\frac{\delta \epsilon_r}{\delta x_\alpha }\vec E_0(\vecr)+\epsilon_r(\vecr)\vec g(\vecr,\vecrp;\vec U_{m,\alpha })\right]-\frac{1}{\epsilon_0}\frac{\delta\rho_e}{\delta x_\alpha }\right\}x_\alpha e^{-i\Omega_{m,\alpha }t} + \mathrm{c.c.} &= 0\\
        \sum_\alpha \nabla\times \vec g(\vecr,\vecrp;\vec U_{m,\alpha })x_\alpha e^{-i\Omega_{m,\alpha }t} + \mathrm{c.c.}&=0,
    \end{align}
\end{subequations}
which yields the equations for the field variation under each eigenmode
\begin{subequations}
    \begin{align}
        \nabla\cdot\left[\epsilon_r(\vecr)\vec g(\vecr,\vecrp;\vec U_{m,\alpha })\right]&=\frac{1}{\epsilon_0}\frac{\delta\rho_e}{\delta x_\alpha }-\nabla\cdot\left[\frac{\delta \epsilon_r}{\delta x_\alpha }\vec E_0(\vecr)\right],\label{eq:Poisson_for_g}\\
        \nabla\times \vec g(\vecr,\vecrp;\vec U_{m,\alpha })&=0.
    \end{align}
\end{subequations}
As evident from these equations, $\vec g$ is equivalent to the electric field generated by two source charges on the right-hand side of \eref{eq:Poisson_for_g}: the first one is a result of displacement of a non-uniform charge density and the second term is a result of deformation of a dielectric medium with non-uniform permittivity. Integration of these terms on the boundaries as well as \eref{eq:E_0_boundary_cond} give rise to three types of boundary sources:

\begin{enumerate}
    \item A discontinuity of the charge density at a boundary gives rise to an equivalent surface charge density for $\vec g$. This surface density can be found by integrating \eref{eq:Poisson_for_g} over a small volume at the boundary. Using the definition \eref{eq:rho_var_def} and the fact that $|\vec U_m|$ is finite, we find this surface density to be
    \begin{equation}\label{eq:free_surface_density}
        \sigma_\alpha = \left(\rho_e^{(2)}-\rho_e^{(1)}\right)\left[\vec U_{m,\alpha }(\vecr)-\vec U_{m,\alpha }(\vecrp)\right]\cdot\vec n
    \end{equation}

    \item The second contribution is the displacement of the surface charges, $\sigma_e$, introduced in \eref{eq:E_0_boundary_cond}, which gives rise to equivalent surface distributions of dipoles for $\vec g$. The corresponding dipole density at $\vecr$ on the boundary is given by
    \begin{equation}\label{eq:free_dipole_density}
        \vec D_{\alpha,e}(\vecr) = \sigma_e(\vecr)\left[\vec U_{m,\alpha }(\vecr)-\vec U_{m,\alpha }(\vecrp)\right].
    \end{equation}
    This density means that an element of area on the boundary has the dipole moment of $\vec D_{\alpha,e}da$.
    \item The third component is due to the discontinuity of the permittivity. In the undeformed geometry wherever that we have a dielectric interface (i.e. discontinuity of the $\epsilon_r$) we have accumulated surface bound charges with the surface density $\sigma_b = (\vec E_0^{(2)}-\vec E_0^{(1)})\cdot \vec n$. Using \eref{eq:Poisson_for_g}, we find that displacement of these charges gives rise to a dipole density
    \begin{equation}\label{eq:bound_dipole_density}
        \vec D_{\alpha,\epsilon}(\vecr) = \left[(\vec E_0^{(2)}-\vec E_0^{(1)})\cdot \vec n\right](\epsilon_r^{(2)} - \epsilon_r^{(1)})\left[\vec U_{m,\alpha }(\vecr)-\vec U_{m,\alpha }(\vecrp)\right].
    \end{equation}
\end{enumerate}

In summary, one can see $\vec g$ as an electric field with the corresponding electric displacement field $\epsilon_0\epsilon_r\vec g$, in the presence a charge density $\delta\rho_e/\delta x_\alpha$ and a permanent polarization vector $\epsilon_0(\delta\epsilon_r/\delta x_\alpha)\vec E_0$ in addition to the boundary sources introduced in Equations \ref{eq:free_surface_density}, \ref{eq:free_dipole_density} and \ref{eq:bound_dipole_density}. In case of homogeneous materials and uniform charge densities, the bulk contributions vanish and we only have to deal with the boundary charges.

\subsection{Green's function method for moving charges}\label{app:Green_method}
In this section we introduce a formal method for calculating the mechanical-dielectric dynamical effects for systems which can be approximated as moving charges interacting with static dielectrics. We start by assuming the undeformed geometry to have an electrostatic Green's function, $\vec{G}(\vecr, \vecrp, \omega)$, in a way that the electric field in the frequency domain is given by
\begin{equation}\label{eq:greens_def}
    \vec{E}(\vecr, \omega) = \Kcoul \int d^3r' \rho_e(\vecrp, \omega)\vec{G}(\vecr, \vecrp, \omega).
\end{equation}
This Green's function is in fact the gradient of the usual scalar Green's function used to compute the scalar potential and it is the solution of equations similar to \eref{eq:Poissons}, with a delta function charge density
\begin{subequations}
    \begin{align}
        \nabla\cdot\left[\epsilon_r(\vecr,\omega)\vec G(\vecr, \vecrp,\omega)\right]&=4\pi\delta(\vecr - \vecrp),\\
        \nabla\times \vec G(\vecr, \vecrp,\omega) &= 0.
    \end{align}
\end{subequations}
The frequency dependence of $\vec G$ comes from the frequency dependence of $\epsilon_r$. Now suppose that the body containing the charges goes under deformation with a displacement field $\vec u (\vecr,\omega)$. When evaluating the electric field, this corresponds to modification of the source coordinates as $\vecrp\rightarrow \vecrp+\vec u (\vecrp,\omega) $. For small oscillations (ie. $|\vec u (\vecrp,\omega)|\ll|\vecrp|$), we can expand the electric field to the first order in $\vec u$. The higher order terms involve more complicated phenomena. For example the second order term results in frequency jitter \cite{yazdanian_dielectric_2008}. We also assume that the charge density has no explicit frequency dependence. One obtains for the electric field
\begin{multline}\label{eq:efield_full}
    \vec{E}(\vecr,\omega) = \Kcoul \int d^3r' \rho_e(\vecrp)\vec{G}(\vecr, \vecrp,\omega)\\
    + \Kcoul \int_{V_r} d^3r' \nabla'\rho_e(\vecrp)\cdot \vec u (\vecrp,\omega)\vec{G}(\vecr, \vecrp,\omega)\\
    + \Kcoul \int_{V_r} d^3r' \rho_e(\vecrp)\nabla'\vec{G}(\vecr, \vecrp, \omega)\cdot \vec u (\vecrp,\omega),
\end{multline}
where $\nabla'$ corresponds to vector derivation with respect to the source coordinate $\vecrp$ and $\nabla'\vec{G}(\vecr, \vecrp,\omega)$ is a $3\times 3$ tensor with elements $\partial G_i(\vecr, \vecrp,\omega)/\partial r'_j$. The first term \eref{eq:efield_full} is the constant electrostatic field and has no dynamical effects. The second term corresponds to direct modification of the charge density due to its non-uniformity and the third term is a result of charge displacement. The integration for these two terms is over the volume of the resonator, $V_r$. For the third term, it can be shown that it is equivalent to the field generated by a polarization density field
\begin{equation}
    \vec P_r(\vecr,\omega)=\rho_e(\vecr)\vec u(\vecr,\omega).
\end{equation}
For uniform charge distribution and simple geometries, this equivalence can be useful for more straightforward calculations. In \appref{app:toy_model}, we have used this equivalence to compute the NCF damping for an infinitesimally thin moving string.

\subsection{Dissipated power and stored energy}
Generally, the volume density of power dissipated in a dielectric is given by $p_d(\vecr,t) = \vec E(\vecr,t)\cdot \partial_t \vec D(\vecr,t)$, where $\vec D(\vecr,t)$ is the electric displacement field. If we plug the field variation from \eref{eq:delta_E} into this relation, we would get different frequency mixing terms which is results in inter-modal damping. In a situation where the modes are sufficiently separated in frequency compared to the damping, these effects are negligible and we can treat each mechanical mode separately with a single frequency varying electric field.

In dielectrics with dispersion and loss the relationship between $\vec D(\vecr,t)$ and $\vec E(\vecr,t)$ is not instantaneous and in the frequency domain is given by $\vec D(\vecr,\omega) = \epsilon_0\epsilon_r(\vecr, \omega)\vec E(\vecr,\omega)$. Here $\epsilon_r(\vecr, \omega)$ is the material's frequency dependent relative permitivity. For harmonic fields with the oscillation frequency $\omega$, expressed by $\vec E(\vecr,t) = \vec E(\vecr,\omega)e^{-i\omega t} + \mathrm{c.c.}$, the period averaged dissipated power density is given by
\begin{equation}\label{eq:p_diss}
    \bar p_d(\vecr) = 2\omega\epsilon_0\epsilon''_r(\vecr, \omega)|\vec E(\vecr,\omega)|^2,
\end{equation}
where $\epsilon''_r(\vecr, \omega) = \mathrm{Im}[\epsilon_r(\vecr, \omega)]$ characterises the material's dielectric loss. The dissipated power by the motion of a mode with displacements similar to \eref{eq:u_rt_expansion} is given by
\begin{equation}\label{eq:P_diss_total_g}
    \bar P_d = \int d^3r 2\Omega_m\epsilon_0\epsilon''_r(\vecr,\Omega_m)|\vec g(\vecr, \vecr;\vec U_{m})|^2|x_m|^2.
\end{equation}
 Note that the field variation at each point is calculated in its own reference frame to account for the relative motion. For this mode, the motion can be reduced to \emph{point-like} particle with the displacement $u(t) = x_m e^{-i\Omega_mt} + \mathrm{c.c.}$. With a perturbative approach, when this particle is subjected to viscous damping force with the form $F_d = -\gamma_\mathrm{NCF}\dot u_m(t)$, the power dissipated by this force, average for each period is given by
 \begin{equation}
     \bar P_d = 2\gamma_\mathrm{NCF}\Omega_m^2|x_m|^2.
 \end{equation}
 Comparing this equation with \eref{eq:P_diss_total_g} yields the result of the main text.
 \begin{equation}\label{eq:gamma_ncf}
    \gamma_\mathrm{NCF} = \int d^3r \,\epsilon_0|\vec g(\vecr, \vecr;\vec U_m(\vecr))|^2\frac{\epsilon''_r(\vecr, \Omega_m)}{\Omega_m}.
\end{equation}

On the other hand, the stored energy within the dielectrics, given by the density $\frac{1}{2}\vec E\cdot \vec D$, also gives rise to an energy term quadratic in the displacements of the mechanical resonator. With a similar approach as above for harmonic motion, the period average of this electrostatic energy is given by
\begin{equation}\label{eq:P_diss_total_g}
    \bar E_\mathrm{es} = \int d^3r \epsilon_0\epsilon'_r(\vecr,\Omega_m)|\vec g(\vecr, \vecr;\vec U_{m})|^2|x_m|^2.
\end{equation}
Referring this energy to an added spring constant $F_\mathrm{es} = -k_\mathrm{es}u_m(t)$ yields
\begin{equation}\label{eq:P_diss_total_g}
    k_\mathrm{es} = \int d^3r \epsilon_0|\vec g(\vecr, \vecr;\vec U_{m})|^2\epsilon'_r(\vecr,\Omega_m).
\end{equation}

\subsection{Forces on the moving bodies}
The dynamical effects discussed above can be found via computing the forces on the mechanical resonator. The perturbation due to the displacements give rise to forces with linear dependency on the position. The in-phase and out-of-phase components of these forces are interpreted as the modified spring constant and the damping coefficient. A general approach for computing the forces, that is consistent with the energy conservation, is via the electrostatic Maxwell's stress tensor, $\textbf{T}$, defined as $T_{ij} = E_iE_j - \delta_{ij}|\vec E|^2$, where $\delta_{ij}$ is the Kronecker's delta. Applying the perturbation $\vec E = \vec E_0 + \vec g(\vecr, \vecr;\vec U_m(\vecr))u_m$ to $\textbf{T}$ yields $\textbf{T} = \textbf{T}_0 + u_m(\delta\textbf{T}/\delta u)$ where
\begin{equation}
    \left(\frac{\delta\textbf{T}}{\delta u}\right)_{ij} = \left(E_{0,i}g_j(\vecr, \vecr;\vec U_m(\vecr)) + E_{0,j}g_i(\vecr, \vecr;\vec U_m(\vecr)) - \delta_{ij}\vec E_0\cdot\vec g(\vecr, \vecr;\vec U_m(\vecr))\right).
\end{equation}
 Note that, similar to before, the field variation at each point is calculated in its own reference frame to account for the relative motion. The force unit volume applied on the resonator is given by $\nabla\cdot \textbf{T}$ (note that this object is a vector) and the point-like force exerted on a certain mode, $\vec U_m$, is given by
 \begin{equation}
     F_u = \int_{V_r}d^3r\vec U_m\cdot \left(\nabla\cdot \textbf{T}\right),
 \end{equation}
 where the integration is over the volume of the resonator. Applying the perturbation field the position dependent force for the mode
 \begin{equation}
     F(u_m) = u_m\int_{V_r}d^3r\vec U_m\cdot\left(\nabla\cdot \frac{\delta\textbf{T}}{\delta u}\right).
 \end{equation}
Referring this force to a spring constant and damping as $F(u_m) = -k_\mathrm{es}u_m-\gamma_\mathrm{NCF}\dot u_m$ gives
\begin{equation}
    k_\mathrm{es} = - \int_{V_r}d^3r \mathrm{Re}\left[\vec U_m\cdot\left(\nabla\cdot \frac{\delta\textbf{T}}{\delta u}\right)\right],
\end{equation}
and
\begin{equation}
    \gamma_\mathrm{NCF} = \frac{1}{\Omega_m} \int_{V_r}d^3r \mathrm{Im}\left[\vec U_m\cdot\left(\nabla\cdot \frac{\delta\textbf{T}}{\delta u}\right)\right].
\end{equation}
Note the inverse frequency scaling appearing again. These forces take into account both the Coulomb forces on the charge as well as the dipole forces on the dielectrics. For situations where we can neglect the dipole forces, there is a major simplification and the force density on the resonator is given by $\rho_e\vec g(\vecr, \vecr;\vec U_m(\vecr))$. The computation in the reference frame of the resonator ensures that the \emph{self-terms} do not appear and there will be no singularities. With this simplification the dynamical effects are given by

\begin{equation}
    k_\mathrm{es} = - \int_{V_r}d^3r \rho_e\mathrm{Re}\left[\vec U_m\cdot\vec g(\vecr, \vecr;\vec U_m(\vecr))\right],
\end{equation}
and
\begin{equation}\label{eq:gamma_NCF_from_force}
    \gamma_\mathrm{NCF} = \frac{1}{\Omega_m} \int_{V_r}d^3r \rho_e\mathrm{Im}\left[\vec U_m\cdot\vec g(\vecr, \vecr;\vec U_m(\vecr))\right].
\end{equation}
These expressions are particularly useful for point-like charges where computing the force is significantly simpler than computing the dissipated power. In the AFM community, this approach is more common as the charge on the cantilever is modeled by a point charge \cite{yazdanian_dielectric_2008}.

\section{Noncontact friction for an infinitesimally thin string}\label{app:toy_model}
\begin{figure}[t!]
\includegraphics{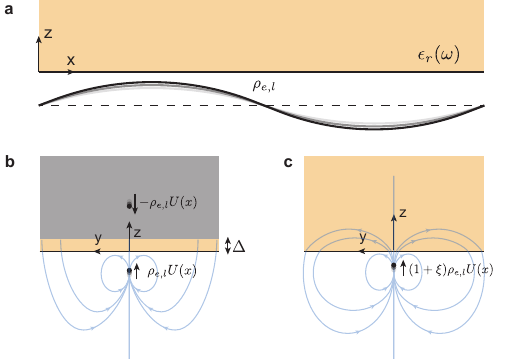}
\caption{\textbf{Noncontact friction for an infinitesimally thin string a} The string is along the x-axis and it oscillates along the z-axis. It located in air, with distance $d$ from a semi-infinite dielectric slab with complex relative permittivity $\epsilon_r(\omega)$. \textbf{b} View of \textbf{a} in the yz plane, showing the motion of the string and the electric field variation (blue streamlines).}
\label{fig:app_toy_model}
\end{figure}
We calculate the NCF damping coefficient for a 1-dimensional tensioned string interacting with a substrate. The schematic is shown in \subfref{fig:app_toy_model}{a}. The string has length $L$ and a linear charge density of $\rho_{e,l}$. The motion of the beam is given by a mode shape $U(x)$ (motion along the z-axis). As shown in \appref{app:Green_method}, for situations where the electric energy dissipated within the volume of the resonator itself can be neglected, $\vec g$ is equivalent to the electric field generated by a polarization density $ \vec P(\vecr) = \rho_e(\vecr)\vec U_m(\vecr)$. Due to the zero volume of the string, to calculate the field variation $\vec g$ we can use this approximation and replace the linear charge density, with a linear polarization density $\vec P_l(x) = \rho_{e,l}U(x)\hat z$. Moreover, for modes where $d\ll |U'(x)|^{-1}$ we can approximately reduce the problem to 2D in the cross-section and compute the field for an infinite line (see \appref{app:3Dto2D}). We consider two different types of substrate and compute the NCF-limited quality factors for each case.

Furthermore, we reproduce the results for the problem a thin string oscillating in front of semi-infinite dielectric slab, as shown in \subfref{fig:app_toy_model}{a} using two methods, other than the one discussed in the one above. We also compute NCF for the situation where the dielectric slab is finite and is placed on a conductor.

\subsection{Noncontact friction due to a conductive substrate with a thin oxide layer}
Even for high resistivity silicon, the characteristic time constant for the charges to decay is given by $\epsilon/\sigma \sim 10^{-8}$ s, rendering silicon as a conductor for our frequency range. Hence we have to treat the silicon substrate as a conductor in our models. For modeling the effect of the native silicon oxide layer, the schematic is shown in \subfref{fig:app_toy_model}{b}. For a thin oxide layer ($\Delta\ll d$), we can neglect the effect of the dielectric layer on the electric field and compute it as if the string was in front of a conductor. Hence, the electric field variation is given by the field generated by the original dipole density $\vec P_l(x)$ in addition to its image $-\vec P_l(x)$ at $z =  \Delta + d$. Moreover, we approximate the field within the layer to be uniform and equal to the field generated by the two dipoles at the boundary, reduced by $\epsilon'_r$. The resulting field within the dielectric only has a $z$ component and is given by
\begin{equation}
    \vec g(\vecr,\omega;U(x)) \approx\frac{\rho_{e,l}}{\pi\epsilon_0\epsilon_r}\frac{\left(y^2-d^2\right)\hat z}{\left(y^2 + d^2\right)^2}U(x).
\end{equation}
To calculate $\gamma_{\mathrm{NCF}}$, the NCF damping for a harmonic mode of the string with frequency $\Omega_m$, we use \eref{eq:gamma_ncf} and integrate $|\vec g|^2$ within the oxide layer
\begin{equation}
    \gamma_{\mathrm{NCF}} = \frac{\rho_{e,l}^2\epsilon_r''}{|\pi\epsilon_0\epsilon_r|^2\Omega_m}\int_{0}^\Delta\int_{-\infty}^{\infty}\int_{0}^{L}dxdydz\frac{\left(y^2-d^2\right)^2}{\left(y^2 + d^2\right)^4}|U(x)|^2.
\end{equation}
The integration along the x-axis is evaluated using the definition of the effective mass. Given that the mode $U_n(x)$ is normalized to its maximum the effective mass is defined as $m_\mathrm{eff} = \rho_{m,l}\int_0^L dx|U(x)|^2$, where $\rho_{m,l}$ is the linear mass density. Carrying out the integration in the yz plain yields
\begin{equation}\label{eq:gamma_ncf_n_beam}
    \gamma_{\mathrm{NCF}} = \frac{m_{\mathrm{eff}}\rho_{e,l}^2\Delta}{4\pi\epsilon_0d^3\rho_{m,l}}\frac{\tan\delta(\Omega_m)}{\epsilon_r'(\Omega_m)\Omega_m}.
\end{equation}
Calculating the damping rate, $\Gamma_{\mathrm{NCF}} = \gamma_{\mathrm{NCF}}/m_\mathrm{eff}$, and the quality factor yields
\begin{equation}
    Q_\mathrm{NCF} = \frac{4\pi\epsilon_0d^3\rho_{m,l}}{\rho_{e,l}^2\Delta}\frac{\epsilon_r'(\Omega_m)\Omega_m^2}{\tan\delta(\Omega_m)}.
\end{equation}
To obtain the numerical estimates for $Q_\mathrm{NCF}$ shown in \subfref{fig:NCF_in_ToyModel}{c}, we have considered the dielectric permittivity and loss tangent of the \ce{SiO2} substrate to be \SI{3.9}{} and $10^{-3}$ respectively \cite{noauthor_dielectric_nodate}. We have also retrieved the designs and the chips from the references \cite{ghadimi_elastic_2018, bereyhi_hierarchical_2022, bereyhi_perimeter_2022}, measured the distance $d$ using an optical microscope and computed the corresponding $Q_\mathrm{NCF}$ for each geometry.

\subsection{Noncontact friction due to a semi-infinite dielectric substrate}
 The schematic for the semi-infinite lossy dielectric substrate is shown in \subfref{fig:app_toy_model}{c}. The $z>0$ space is filled by the dielectric with relative permittivity $\epsilon_r(\omega)$. Using an elementary method of image charges, we calculate the field within the dielectric ($z>0$) by replacing $\vec P_l(x)$ with $(1+\xi)\vec P_l(x)$.  Following these steps, we find the electric field variation for $z>0$ as
\begin{equation}\label{eq:g_string_dielectric}
    \vec g(\vecr,\omega;U(x)) =\frac{(1+\xi)\rho_{e,l}}{2\pi\epsilon_0\epsilon_r}\times
    \frac{2y(z+d)\hat y+\left[(z+d)^2-y^2\right]\hat z}{\left[y^2 + (z+d)^2\right]^2}U(x).
\end{equation}
To calculate $\gamma_{\mathrm{NCF},n}$, the NCF damping for a harmonic mode of the string $U_n$, we use \eref{eq:gamma_ncf}.
\begin{equation}
    \gamma_{\mathrm{NCF},n} = \frac{|1+\xi|^2\rho_{e,l}^2}{4\pi^2\epsilon_0|\epsilon_r|^2}\frac{\epsilon''_r}{\Omega_n}\times\int_{0}^\infty\int_{-\infty}^{\infty}\int_{0}^{L} dxdydz\frac{|U_n(x)|^2}{\left[y^2+(z+d)^2\right]^2}.
\end{equation}
Carrying out the integrations similar to the previous case, one obtains
\begin{equation}\label{eq:gamma_ncf_n_beam}
    \gamma_{\mathrm{NCF},n} = \frac{m_{\mathrm{eff},n}\rho_{e,l}^2}{8\pi\epsilon_0d^2\rho_{m,l}}\frac{\mathrm{Im}[\xi(\Omega_n)]}{\Omega_n},
\end{equation}
where we have used the relation $\mathrm{Im}[\xi(\Omega_n)] \approx \frac{2\epsilon_r''}{(1+\epsilon_r')^2}$ with the assumption $\epsilon_r''\ll\epsilon_r'$. Calculating the damping rate, $\Gamma_{\mathrm{NCF},n} = \gamma_{\mathrm{NCF},n}/m_{\mathrm{eff},n}$, and the quality factor yields \eref{eq:Q_ncf_n}.

\begin{figure}[t!]
\includegraphics{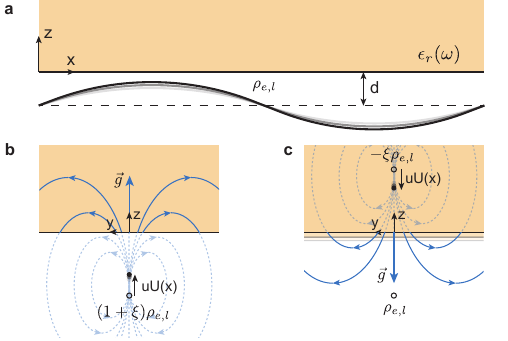}
\caption{\textbf{Other methods for computing NCF for a thin string a} The string is along the x-axis and it oscillates along the z-axis. It located in air, with distance $d$ from a semi-infinite dielectric slab with complex relative permittivity $\epsilon_r(\omega)$. \textbf{b} View of \textbf{a} in the yz plane, showing the motion of the string and the electric field variation (blue streamlines) in $z>0$ in the reference frame of the dielectric. \textbf{c} Same as \textbf{b}, but showing the field variation in $z<0$ in the reference frame of the string.}
\label{fig:app_toy_model}
\end{figure}
\subsection{Semi-infinite dielectric slab: explicit approach}
We compute the NCF damping coefficient by the explicit method of taking the derivative of the stationary field with respect to the amplitude of motion. The schematic is shown in \subfref{fig:app_toy_model}{b}. To find the field in $z>0$, using the technique of image charges we can simply replace the charge density of the beam with $(1+\xi)\rho_{e,l}$ (with $\xi = \frac{\epsilon_r - 1}{\epsilon_r + 1}$). When $d\ll L$, the static field is approximately given by the field from an infinitely long line of charge.
\begin{equation}
    \vec E_0(\vecr;d) = \frac{(1+\xi)\rho_{e,l}}{2\pi\epsilon_0\epsilon_r}\frac{y\hat y + (z+d)\hat z}{y^2 + (z+d)^2}
\end{equation}
To calculate the field variation $\vec g(\vecr)$ under the deformations given by a mode shape $U(x)$ (motion along the z-axis), in the frame of reference of the dielectric slab, we consider a displacement with amplitude $q$ for the line of charge. For modes where $d\ll |U'(x)|^{-1}$ we can approximately reduce the problem to 2D in the cross-section. As illustrated in \subfref{fig:app_toy_model}{b}, we calculate the field in coordinate $x$ from a line of charge  that is displaced by $uU(x)$. The field variation is given by
\begin{equation}
    \vec g(\vecr;U(x),\omega) =\frac{\partial}{\partial u}\left[ \vec E_0(\vecr;d-uU(x))\right]\bigg|_{u=0}=\frac{(1+\xi)\rho_{e,l}}{2\pi\epsilon_0\epsilon_r}\frac{2y(z+d)\hat y+\left[(z+d)^2-y^2\right]\hat z}{\left[y^2 + (z+d)^2\right]^2}U(x).
\end{equation}
Which is the same result as \eref{eq:g_string_dielectric}.
\subsection{Semi-infinite dielectric slab: out-of-phase force}
To compute the damping coefficient using the out-of-phase force, we have to compute $\vec g$ in the frame of reference of the string. As shown in \subfref{fig:app_toy_model}{c}, in this frame the string is stationary and the dielectric slab oscillates. As discussed in sectionn \ref{app:detailed_theory}, the source for $\vec g$ in this situation is the moving bound surface density given by the discontinuity of $\vec E_0$ at $z=0$. But according to the image charges technique, the field made by these charges in $z<0$ is equivalent to the one generated by an image charge with density $\xi\rho_{e,l}$ at $z=d$. Hence, similar as the previous section, for displacement of $uU(x)$, the electric field variation in $z<0$ is given by
\begin{equation}
    \vec g(\vecr;U(x),\omega) =\frac{\partial}{\partial u}\left[ \frac{-\xi\rho_{e,l}}{2\pi\epsilon_0}  \frac{y\hat y + (z-d+uU(x))\hat z}{y^2 + (z-d+uU(x))^2}\right]\bigg|_{u=0}=
    \frac{\xi\rho_{e,l}}{2\pi\epsilon_0}\frac{2y(z-d)\hat y+\left[(z-d)^2-y^2\right]\hat z}{\left[y^2 + (z-d)^2\right]^2}U(x).
\end{equation}
Using \eref{eq:gamma_NCF_from_force}, the NCF damping coefficient is given by
\begin{equation}
    \gamma_\mathrm{NCF} = \frac{1}{\Omega_n}\int_0^Ldx\rho_{e,l}\mathrm{Im}\left[U(x)\hat z\cdot\vec g(x,0,-d)\right],
\end{equation}
which yields
\begin{equation}
    \gamma_\mathrm{NCF} = \frac{\rho_{e,l}^2}{2\pi\epsilon_0(2d)^2}\frac{\mathrm{Im}\left[\xi \right]}{\Omega_n}\int_0^L|U(x)|^2dx.
\end{equation}
Using the definition of the effective mass $m_\mathrm{eff}=\rho_{m,l}\int_0^L|U(x)|^2dx$ we obtain the same result as before.

\subsection{A finite dielectric slab on a conductor}
\begin{figure}[t!]
\includegraphics{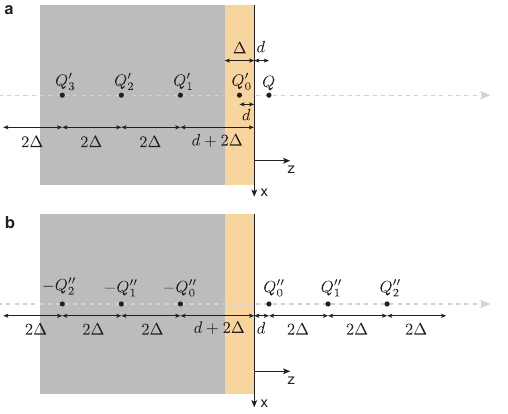}
\caption{\textbf{Image charges for a finite dielectric slab on a conductor a} The image charges for the solution in $z>0$. \textbf{b} Image charges for the solution in $-\Delta<z<0$/.}
\label{fig:app_toy_mode_conductor}
\end{figure}
The schematic of this problem is shown in \fref{fig:app_toy_mode_conductor}. A dielectric layer with the complex permittivity $\epsilon_r$ and thickness $\Delta$ is located on top of a conductor and the string has a distance $d$ from the dielectric surface. To solve this problem, we first find the appropriate image charges configurations for a point charge $Q$ at $z=d$. These configurations for finding the electric field in $-\Delta<z<0$ and $z>\Delta$ are shown in \subfref{fig:app_toy_mode_conductor}{a} and \subfref{fig:app_toy_mode_conductor}{b} respectively. One can easily show that the boundary conditions at $z=0$ and $z=-\Delta$ are satisfied with the following values for the image charges. With $Q'_i = \alpha_iQ$ and $Q''_i = \beta_iQ$ we have
\begin{align}
    \alpha_0 &= -\xi,\\
    \alpha_i &= \frac{(1-\xi^2)}{\xi}(-\xi)^i \quad (i=1,2,...),\\
    \beta_i &= (1+\xi)(-\xi)^i, \quad (i=0,1,...),
\end{align}
where $\xi = \frac{\epsilon_r - 1}{\epsilon_r + 1}$. To find the NCF damping coefficient we use the force calculation method as it is much easier. Similar to the previous section, to find the force on the string at $z = d$ due to the displacements, we place the lines dipole densities with dipole moments $\vec P_l^i = \alpha_i \rho_{e,l}U(x)\hat z$ at the locations of $Q'_i$ in \subfref{fig:app_toy_mode_conductor}{a}. A calculation similar to the previous section yields
\begin{equation}
    \gamma_\mathrm{NCF} = \frac{m_\mathrm{eff}\rho_{e,l}^2}{8\pi\epsilon_0\rho_{m,l}d^2}\mathrm{Im}\left[\xi + \frac{1-\xi^2}{\xi}\sum_{i=1}^\infty\frac{(-\xi)^i}{\left(1 + i\frac{\Delta}{d}\right)^2}\right].
\end{equation}
Similar to the approach in \cite{kuehn_noncontact_2006}, we evaluate the infinite summation by replacing the denominators using the identity $b^{-2} = \int_0^\infty dv ve^{-bv}$
\begin{equation}
    \gamma_\mathrm{NCF} = \frac{m_\mathrm{eff}\rho_{e,l}^2}{8\pi\epsilon_0\rho_{m,l}d^2}\frac{\mathrm{Im}\left[\xi \right]}{\Omega_n}\left(1-r(\frac{\Delta}{d}, \xi)\right),
\end{equation}
where
\begin{equation}\label{eq:r_correction}
    r(\frac{\Delta}{d}, \xi) = \frac{\mathrm{Im}\left[(\xi^2-1)\int_0^\infty dv\frac{ve^{-(1+\frac{\Delta}{d})v}}{1 + \xi e^{\frac{\Delta}{d}v}}\right]}{\mathrm{Im}\left[\xi \right]}.
\end{equation}
For \ce{SiO2} with $\epsilon_r = 3.9(1 + \tan\delta)$, $r$ is shown in \fref{fig:app_toy_mode_conductor_r} as a function of $\Delta/d$, for different values of the loss tangent. As one can see, the correction due to the presence of the conductive substrate is smaller for thicker oxide layers and the model converges to the semi-infinite dielectric substrate. For the thicknesses used in our experiments this correction is less than 10\%.
\begin{figure}
\includegraphics{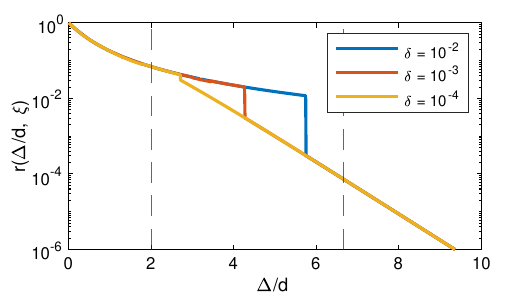}
\caption{\textbf{Correction to the NCF damping coefficient due to the conductive substrate.} The correction factor, $r(\frac{\Delta}{d}, \xi)$, defined in \eref{eq:r_correction} as a function of $\Delta/d$. Values are computed for \ce{SiO2} with the permittivity $\epsilon_r = 3.9(1 + \tan\delta)$, for three different values of $\delta$.}
\label{fig:app_toy_mode_conductor_r}
\end{figure}
\section{General 3D to 2D reduction}\label{app:3Dto2D}
In this we introduce the mathematical details of the procedure for reducing a problem with translational symmetry from 3D to 2D. This procedure is useful both in analytical solutions and FEM simulations. Consider a geometry where $\vec E_0$ does not depend on one of the coordinates (say, $z$). Moreover, the mode shape also varies slowly with $z$ and can be written as $\vec U(x,y,z) = \vec U_\perp(x,y)U_\parallel(z)$. We define the mode cross section as
\begin{equation}
    S_\perp = \int dxdy|\vec U_\perp(x,y)|^2.
\end{equation}
One can easily show that the effective mass of the mode is given by $m_\mathrm{eff} = \rho_mS_\perp\int dz U_\parallel(z)^2$. For such a geometry, the field variation $\vec g$ can be approximated as
\begin{equation}
    \vec g(x, y,z) = \vec g_\perp(x,y;\vec U_\perp(x,y))U_\parallel(z).
\end{equation}
Here $\vec g_\perp$ is the transverse counterpart of $\vec g$. With these definitions, the NCF damping coefficient can be written as
\begin{equation}
    \gamma_\mathrm{NCF} = \frac{\epsilon_0}{\Omega_m}\int dxdy\epsilon_r''(x,y)|\vec g_\perp(x,y)|^2\int dzU_\parallel(z)^2,
\end{equation}
which allows us to reduce the computations for problems with this symmetry. For the case of the uniform beams above dielectric surfaces, since the length of the dielectric surface coincides with the length of the resonator we can further simplify
\begin{equation}
    \gamma_\mathrm{NCF} = \frac{\epsilon_0m_\mathrm{eff}}{\Omega_m\rho_{m,l}}\int dxdy\epsilon_r''(x,y)|\vec g_\perp(x,y)|^2,
\end{equation}
where $\rho_{m,l} = \rho_mS_\perp$. These relations are used for the NCF calculations based on the FEM simulation of the cross section of the devices.

\section{Dissipation-dilution limit to force sensitivity}\label{app:dissipation_dilution_limit_to_S_F}
For a string with thickness $h$, stress $\sigma$ and mass density $\rho_m$, the ultimate \emph{clampless} limit for the Q at a given frequency $\Omega$ from dissipation-dilution is given by \cite{fedorov_generalized_2019}
\begin{equation}\label{eq:Q_clampless_limit}
    \frac{Q_m^\mathrm{cl}}{Q_\mathrm{int}} \leq \frac{12\sigma^2}{\rho_mEh^2\Omega^2},
\end{equation}
where $E$ is the Young's modulus. To obtain a limit for the force sensitivity at a given string width, $w$, we set the effective mass to the value for the fundamental mode of a uniform string with frequency $\Omega$, $m_\mathrm{eff}^\mathrm{cl} = \frac{\rho_mwh}{2\Omega}\sqrt{\frac{\sigma}{\rho_m}}$. This is the minimal value for a given thickness and width of the string and for soft-clamped modes this value is up to a factor 2 larger. Together, the limit of force sensitivity is given by
\begin{equation}
    S_F^\mathrm{th, cl} = 4k_BT\frac{Eh^3w}{24}\left(\frac{\rho_m}{\sigma}\right)^{\frac{3}{2}}\Omega^2.
\end{equation}

\section{Ruling out other damping mechanisms}\label{app:rule_out_others}
We discuss three mechanisms that could explain a distance-dependent damping in both the uniform strings and the nano-optomechanical system. We show that none of them can fully explain our observations.

\begin{figure}[t!]
\includegraphics{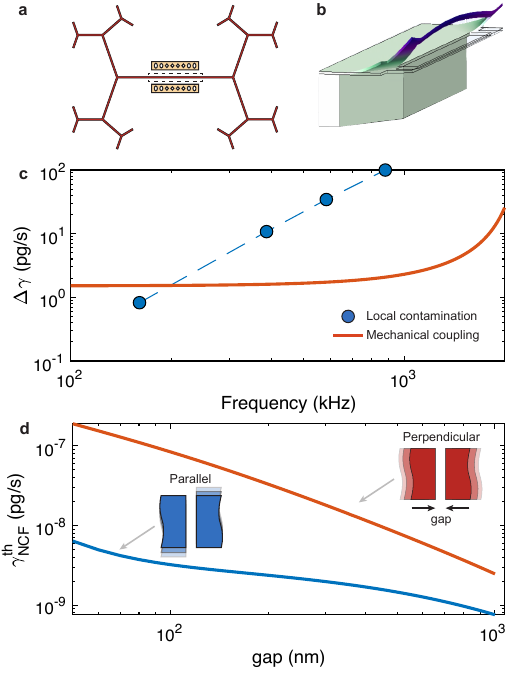}
\caption{\textbf{Ruling out other possible damping mechanisms. a} Schematic of the nano-optomechanical device. The area marked by the dashed box is prone to local reduction of the intrinsic Q. \textbf{b} FEM simulation of the fundamental mechanical mode of the PhC cavity at \SI{2.1}{\mega\hertz}. \textbf{c} Additional damping coefficients caused by local contamination (blue circles) and mechanical coupling to the PhC cavity (orange line) versus frequency. \textbf{d} Estimation of the fundamental thermal-electrodynamic damping coefficient for two silicon nitride semi-infinite bodies in parallel (blue) and perpendicular (orange) relative motion versus their distance. }
\label{fig:rule_out}
\end{figure}
\subsection{Gas and squeeze-film damping}
Although the samples are tested under high vacuum ($P<10^{-8}$ mbar), the small gaps can result in enhancement of gas damping due to squeeze-film damping. However, squeeze-film gas damping results in a frequency-independent damping rate \cite{martin_damping_2008, bao_squeeze_2007}, different from our observations.

\subsection{Local surface contamination}
Local surface contamination means reduction of the intrinsic mechanical $Q$ in a local area on the nano-beam. For example this could be a result of the etching of the PhCs in the vicinity of the beam. Such a mechanism could in principle explain the different behavior of modes with different parities. To study this effect, we use a modified version of the FEM simulations used for estimating the dilution factor for tensioned and high-aspect-ratio resonators \cite{fedorov_mechanical_2021}. These simulations are based on evaluating the kinetic and bending energies, $W_\mathrm{kin}$ and $W_\mathrm{bend}$, for a given mode and then computing the quality factor as $Q = W_\mathrm{kin} / W_\mathrm{bend}$. For a 2D resonator (in xy) plane and an out-of-plane mode with frequency $\Omega$ and mode shape $U(x,y)$ these energies are given by
\begin{equation}
    W_\mathrm{kin} = \frac{1}{2}\rho_mh\Omega^2\int_Sdxdy U(x,y)^2,
\end{equation}
\begin{equation}
    W_\mathrm{bend} = \frac{Eh^3}{24(1-\nu^2)}\int_Sdxdy\phi(x,y)\times\left[(U_{xx}+U_{yy})^2 + 2(1-\nu)(U_{xy}^2-U_{xx}U_{yy})\right],
\end{equation}
where $\rho_m$ is the mass density, $h$ is the thickness, $E$ is the Young's modulus and $\nu$ is the Poisson's ratio of the resonator. The integrations are carried over $S$, the area of the resonator. $\phi(x,y)$ defines the local mechanical loss angle of the resonator where in the usual cases, it is constant and given by the inverse of the material's intrinsic quality factor $\phi(x,y) = Q_\mathrm{int}^{-1}$.

For the resonator design used to take the data in \subfref{fig:NCF_in_OM}{f}, we use this FEM simulation to estimate the effect of local contamination. In a region on the resonator nearby the PhC's location (see \subfref{fig:rule_out}{a}), we set $\phi$ higher than the usual value so that we reproduce the observed reduced quality factor. We compute the Q for the first few OP modes and compute the added damping coefficient compared to the fully \emph{clean} case (similar to \eref{eq:gamma_NCF_exp}). We show these values for the modes with an anit-node in the center in \subfref{fig:rule_out}{c}. We observe an increasing frequency dependence for the added damping coefficient which is unable to explain the experimental observations. 

\subsection{Mechanical coupling to the photonic crystal}
Another mechanism that can introduce a distance-dependent loss in the nano-optomechanical system is coupling of the high-Q mode to the low-Q mechanical modes of the PhC cavity \cite{guo_integrated_2022}. \subfref{fig:rule_out}{b} shows an FEM simulation of the fundamental mode of the PhC cavity at \SI{2.1}{\mega\hertz}. We also observe this in the experiments. This mode can couple to the modes of the binary-tree resonator through multiple mechanisms such as electrostatic or Van der Waals forces. Generally, for two mechanical modes with frequencies $\Omega_{1(2)}$, damping rates $\Gamma_{1(2)}$ and effective masses $m_{1(2)}$, that are coupled by a spring constant $k_c$, the additional damping for mode 1 is given by
\begin{equation}
    \delta\Gamma_1 = \frac{k_c^2}{m_1m_2}\frac{\Gamma_2}{(\Omega_2^2-\Omega_1^2)^2+\Gamma_2^2\Omega_1^2 },
\end{equation}
having a resonance-like frequency dependence. For the measured parameters for the two mechanical modes, we set $k_c$ to a value to reproduce the additional damping for the fundamental mode of the binary-tree resonator at \SI{160}{\kilo\hertz}. The result is shown in \subfref{fig:rule_out}{c}, clearly inconsistent with our observed frequency scaling.

\subsection{Fundamental thermal-electrodynamic damping}
In the absence of static charges, the finite thermal fluctuations of the electromagnetic field gives rise to the ultimate form of NCF \cite{volokitin_near-field_2007}. This mechanism that often goes by different names (e.g. vacuum friction. Van der Waals friction, Casimir friction) is a direct consequence of the fluctuation-dissipation theorem for the electromagnetic field in the presence of lossy dielectrics. When two dielectrics are in relative motion, the retardation of the fluctuating electromagnetic field in conjunction with the force exerted on the dielectrics through Maxwell's stress tensor, gives rise to friction forces. Estimating the damping force for arbitrary geometries is complicated, but there is an analytical solution for the case of two semi-infinite bodies \cite{volokitin_noncontact_2003}. We estimate the damping coefficients, $\gamma_\mathrm{NCF}^\mathrm{th}$, for silicon nitride bodies for both parallel and perpendicular relative motions. As shown in \subfref{fig:rule_out}{d}, even for distances of tens of nanometers, $\gamma_\mathrm{NCF}^\mathrm{th}$ is orders of magnitude smaller than our observed values. Moreover, it has been shown that for a mechanical oscillator, this form of damping does not depend on the harmonic potential of the oscillator (i.e. $\gamma_\mathrm{NCF}^\mathrm{th}$ is frequency independent) \cite{zurita-sanchez_friction_2004}.

\section{Finite element model for computation of NCF coefficient }\label{app:FEM_NCF}
\begin{figure*}
    \includegraphics[]{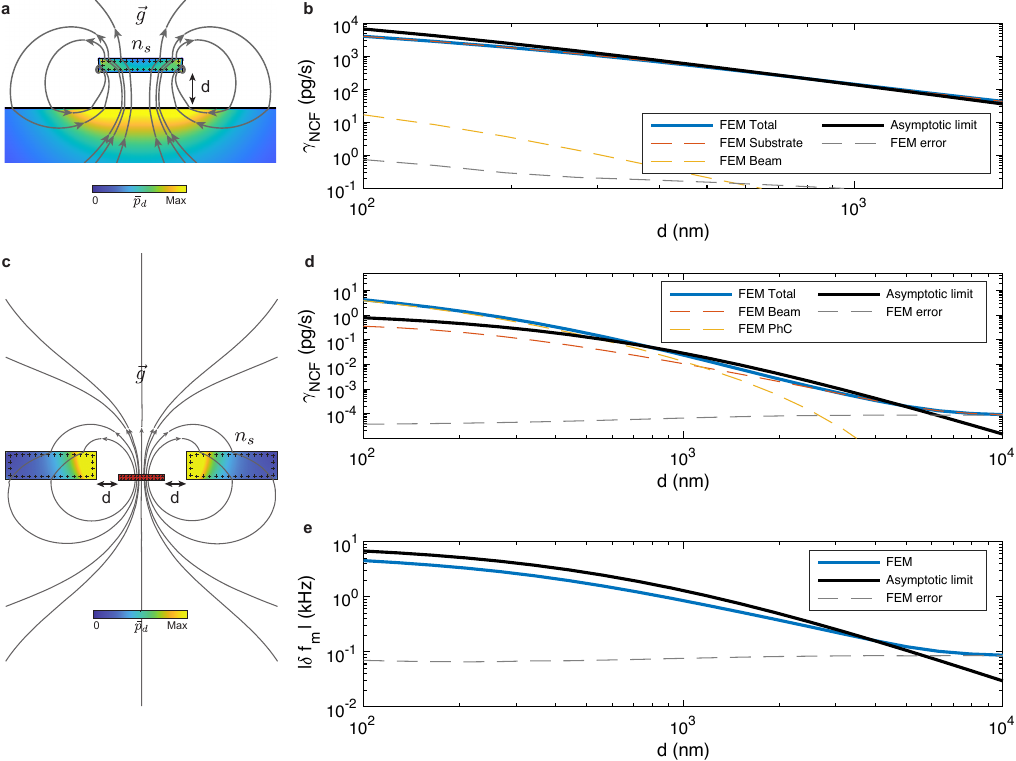}
    \caption{\textbf{Asymptotic behavior of the FEM models a} FEM simulation of the rectangular string above a \ce{SiO2} substrate. All the surfaces of the string are charged with a uniform density. The simulation volume is bounded by sufficiently far zero-potential boundaries (not shown). \textbf{b} NCF damping coefficient and its constituent components vs. gap size, obtained from the simulation in \textbf{a} using \eref{eq:app_ncf_1d_beam_components}. Dashed gray line: the numerical error of the FEM simulation, equal to the value obtained for $\gamma_\mathrm{NCF}^\mathrm{beam}$ when the dielectric substrate is removed. \textbf{c} FEM simulation of the nano-optomechanical system. All surfaces are charged with a uniform charge density. \textbf{d} Same components as \textbf{b}, computed for the simulation in \textbf{c}. \textbf{e} Electrostatic frequency of the string obtained for the simulation in \textbf{c} and using \eref{eq:delta_omega_es}.}
    \label{fig:appendix_FEM_benchmark}
\end{figure*}
Our method is based on the definition of the field variation noted in the main text
\begin{equation}
    \vec g = \frac{\partial \vec E_0}{\partial x},
\end{equation}
where $\vec E_0$ is the electric for the undeformed geometry and $x$ is the amplitude of motion of for a particular mechanical mode. The finite element simulation computes the electric field for the two cases of undisplaced and marginally displaced geometry and then calculates the field variation using a numerical derivative. The steps of the FEM simulation are summarized in the following way:
\begin{enumerate}
    \item The mechanical mode shape is simulated (without any charges or electrostatic interaction).
    \item A charge distribution is defined on the resonator.
    \item The electric field is simulated at two positions of the mechanical resonator: first with no displacement and second displaced by the amplitude $\Delta x$ on the mode shape of interest.
    \item The field variation, $\vec g$ is computed numerically based on the two simulations $\vec g = \frac{\vec E(\Delta x) - \vec E(\Delta x = 0)}{\Delta x}$.
    \item NCF damping is computed using the volume integral formula
    \begin{equation}\label{eq:vol_integral}
        \gamma_\mathrm{NCF} = \frac{\epsilon_0}{\Omega_m}\int d^3r\epsilon_r''|\vec g|^2
    \end{equation}
\end{enumerate}
To account for the relative motion and to calculate the loss within the moving bodies, the simulation has to be done on a mesh that deforms alongside the geometry. We use the \emph{moving mesh} feature in COMSOL Multiphysics. 

In most physical systems, the geometry is composed of discrete bodies, each with a constant dielectric constant. In such systems we can rewrite \eref{eq:vol_integral} as
\begin{equation}
    \gamma_\mathrm{NCF} = \frac{1}{\Omega_m}\sum_i2\tan\delta_iW_i
\end{equation}
where
\begin{equation}
    W_i = \frac{1}{2}\epsilon_0\epsilon_{r,i}\int_{V_i}d^3r |\vec g|^2,
\end{equation}
is the electrostatic energy (generated by $\vec g$) within the body with index $i$, volume $V_i$, relative permittivity $\epsilon_{r,i}$ and dielectric loss tangent of $\tan\delta_i$.

Both physical systems that we have discussed in this paper, to a good approximation, have the 1D translational symmetry. The numerical calculations can be done in the 2D cross section of the systems and using the method of Appendix \ref{app:3Dto2D} NCF damping coefficient can be written as
\begin{equation}
    \gamma_\mathrm{NCF} = \frac{\int_{L_\mathrm{int}} dz U_\parallel(z)^2}{\Omega_m}\sum_i2\tan\delta_iW_{\perp,i},
\end{equation}
where
\begin{equation}
    W_{\perp,i} = \frac{1}{2}\epsilon_0\epsilon_{r,i}\int_{S_i}d^2r |\vec g_\perp|^2,
\end{equation}
is the integration of $g_\perp$ over the cross section of the $i$-th body and $L_\mathrm{int}$ is the length along which the mechanical mode interacts with the dielectrics.

\subsection{Uniform beams above a dielectric surface}

As discussed in Appendix \ref{app:3Dto2D}, for this system $L_\mathrm{int} = L_\mathrm{beam}$ and $\gamma_\mathrm{NCF}$ is given by the sum of two components: dissipation within the \ce{SiO2} substrate and the dissipation within the \ce{Si3N4} beam itself.
\begin{equation}\label{eq:app_ncf_1d_beam_components}
    \gamma_\mathrm{NCF} = \gamma_\mathrm{NCF}^\mathrm{beam} + \gamma_\mathrm{NCF}^\mathrm{substrate},
\end{equation}
where
\begin{equation}
    \gamma_\mathrm{NCF}^\mathrm{i} = \frac{m_\mathrm{eff}}{\rho_{m,l}\Omega_m}2W_i\tan\delta_i.
\end{equation}
As shown in \subfref{fig:appendix_FEM_benchmark}{a}, we simulate a \ce{Si3N4} $100\times 400$ nm beam, uniformly charged on its surface, suspended above a \ce{SiO2} substrate. We show $\gamma_\mathrm{NCF}$, as well as its two components for different gaps in \subfref{fig:appendix_FEM_benchmark}{b}. To benchmark this simulation, we compare its asymptotic behavior for large gaps, to an infinitesimally thin beam with equal linear charge density. We also show the analytical value for the thin beam in \subfref{fig:appendix_FEM_benchmark}{b} and observe good convergence. We note that the convergence can be improved by enlarging the simulation region, which is resource-consuming for large gaps.

In the main text in Section II, to obtain the surface charge densities we have the effective linear density obtained from the fits. The relation between this effective density and the surface density for uniform distribution is obtained by comparing \eref{eq:app_ncf_1d_beam_components} and the analytical result of the thin string which yields,
\begin{equation}
    \rho_{e,s}^2 = \frac{{\rho_{e,l}^\mathrm{eff}}^2\epsilon'_{r,\mathrm{sub}}}{8\pi\epsilon_0d^2(1+\epsilon'_{r,\mathrm{sub}})^2}
    \frac{\tan\delta_\mathrm{sub}}{\tan\delta_\mathrm{sub}W_\mathrm{sub} + \tan\delta_\mathrm{beam}W_\mathrm{beam}},
\end{equation}
where in the FEM simulations for computing $W_i$ we have assumed $\rho_{e,s} = 1$.
\subsection{Integrated nano-optomechanical system}
\begin{figure*}
    \includegraphics[]{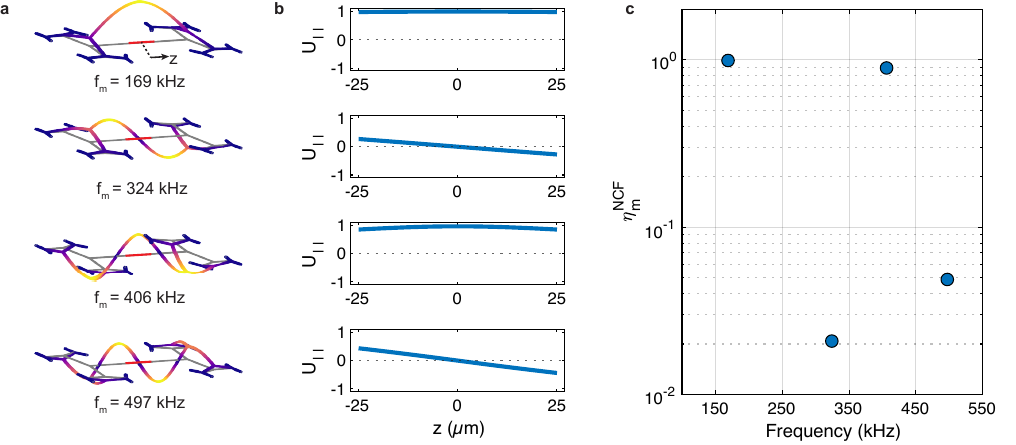}
    \caption{\textbf{Contribution of different modes of the binary-tree resonators to NCF a} FEM simulations of the mode shapes of the first 4 OP modes of the resonator used in the main text to study the dependence on the mode shape. \textbf{b} Mode shape along the red line marked in \textbf{a} for different modes. \textbf{c} Mode contribution to NCF, $\eta_m^\mathrm{NCF}$, computed from the mode shapes in \textbf{b} using \eref{eq:eta_m_int}.}
    \label{fig:appendix_HYB_NCF_overlap}
\end{figure*}
For this system, neglecting the holes of the photonic crystals, the interaction length equals the length of the PhC, $L_\mathrm{int} = L_\mathrm{PhC}$. To use the approximations of Appendix \ref{app:3Dto2D}, we need two conditions: (i) to have $\vec E_0$ (i.e. the electric field at the equilibrium position of the resonator) to be $z$-independent and (ii) to have the mechanical mode to be separable to transverse and longitudinal components. Since the gap in these devices ($d<$\SI{600}{\nano\meter}) is much smaller than the interaction length ($L_\mathbf{int} = L_\mathrm{PhC} \sim$\SI{50}{\micro\meter}) the first condition is met within the interaction region. In the binary-tree resonators used in this paper, the flexural mechanical modes within the interaction length are nearly one-dimensional functions. Hence the second condition is also met and we can reduce the problem to 2D and write the NCF damping coefficient as
\begin{equation}
    \gamma_\mathrm{NCF} = \frac{\eta_m^\mathrm{int}}{\Omega_m}L_\mathrm{cav}\sum_i2\tan\delta_iW_{\perp,i},
\end{equation}
where
\begin{equation}\label{eq:eta_m_int}
    \eta_m^\mathrm{int} = \frac{1}{L_\mathrm{cav}}\int_{L_\mathrm{cav}}dz|U_\parallel(z)|^2
\end{equation}
is the contribution of the mechanical mode (indexed with $m$) to the NCF dissipation within the cavity. $\eta_m^\mathrm{int}$ is computed from the mechanical mode simulation and is between 0 and 1. An example of computation of $\eta_m^\mathrm{int}$ is shown in \fref{fig:appendix_HYB_NCF_overlap}. Similarly to the uniform beams, we can write $\gamma_\mathrm{NCF}$ for this system as 
\begin{equation}
    \gamma_\mathrm{NCF} = \gamma_\mathrm{NCF}^\mathrm{beam} + \gamma_\mathrm{NCF}^\mathrm{PhC},
\end{equation}
corresponding to the dissipation within the beam and the PhC cavities. These components are computed from our FEM model shown in \subfref{fig:appendix_FEM_benchmark}{c}. Similar to the main text, we have assumed uniform charge distribution on all the surfaces. We show the total NCF damping coefficient and its constituent components in \subfref{fig:appendix_FEM_benchmark}{b}. To compute the asymptotic limit at far distances, we model the PhC cavities as cylinders exposed the uniform electric generated by the beam where the field within the dielectric cylinder acquires a factor $\frac{2}{1+\epsilon_r^{-1}}$. The deviation of the FEM value from the asymptotic limit is due to the finite size of the simulation space which leads to an error shown in the same figure. This deviation can be reduced by augmenting the simulation space.

Our FEM simulation can be used to compute the frequency shift in the integrated samples induced by the electrostatic force. When the volume of the resonator is negligible, we can compute the spring constant using by computing the force exerted on the charges by the field variation. In this case, for a small displacement of the mode $\vec U_m$ with the amplitude $x$, the resonator is exposed to a force volume density of $\vec f_r = \rho_e\vec gx$. The point-like equivalent of this force for this mode is given by $F_r = \int d^3r\vec U_m\cdot \vec f_r$ which gives $F_r = x\int d^3r\rho_e\vec U_m\cdot \vec g$. Now we can define the electrostatic-induced spring constant as
\begin{equation}
    k_\mathrm{es} = \int d^3r\rho_e\vec U_m\cdot \vec g.
\end{equation}
When reducing to 2D, we obtain
\begin{equation}
    k_\mathrm{es} = -\int_{L_\mathrm{cav}}dz|U_\parallel(z)|^2\int_{S_r}d^2r\rho_e\vec U_\perp\cdot \vec g,
\end{equation}
where we have neglected the force exerted on the resonator outside the interaction region. The mode shape in 2D is only a unit vector $\vec e_\perp$. Using the simulated $\vec g_\perp$ with constant surface density $\rho_{e,s}$ on the resonator, we can compute the frequency shiftconstant 
\begin{equation}\label{eq:delta_omega_es}
    \delta\Omega_m = -\rho_{e,s}\frac{\eta_m^\mathrm{int}}{2m_\mathrm{eff}\Omega_m^2}L_\mathrm{cav}\int_{\partial S_i}dr\vec e_\perp\cdot \vec g_\perp.
\end{equation}
This quantity is shown in \subfref{fig:appendix_FEM_benchmark}{e}. The asymptotic limit for this case is when the PhCs and string are so far that they can be approximated as lines of charge for which we have an analytical result
\begin{equation}
    \delta\Omega_m = -\frac{\eta_m^\mathrm{int}}{2m_\mathrm{eff}\Omega_m}\frac{\rho_{e,l}^\mathrm{beam}\rho_{e,l}^\mathrm{PhC}}{\pi\epsilon_0d^2},
\end{equation}
where $\rho_{e,l}^i$ is the charge per unit length for the string and PhCs. This result is also shown in \subfref{fig:appendix_FEM_benchmark}{e}, showing a good agreement with the FEM result. The agreement can be improved by augmenting the simulation volume.

\section{Supplementary data for the width dependence}\label{app:width_dependence}
\begin{figure*}
\includegraphics{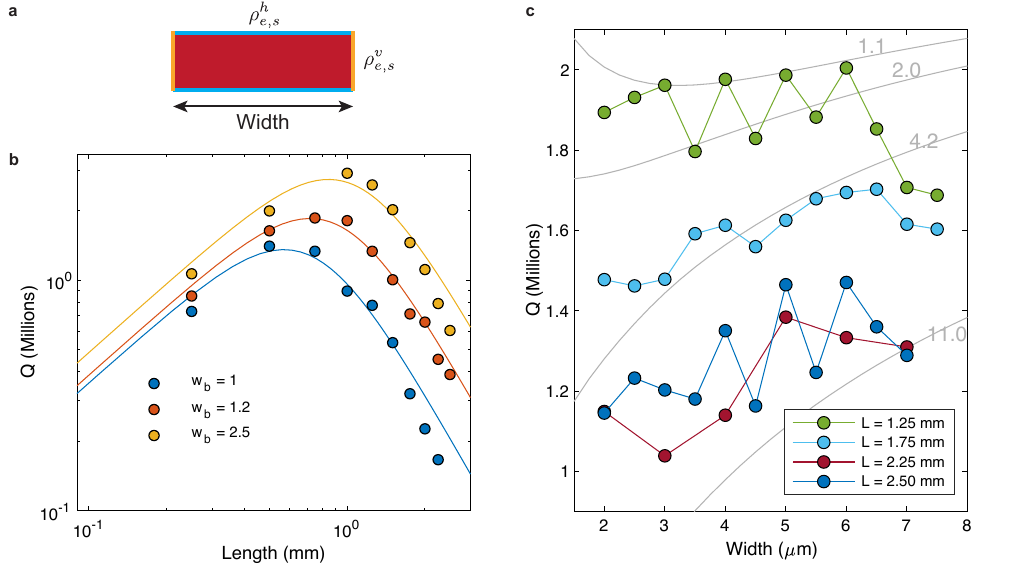}
\caption{\textbf{Dependence of the Q of the uniform strings above a substrate on their width a} Cross section of a string showing different surface charge densities on the horizontal (blue) and vertical (yellow) surfaces. \textbf{b} Q of the fundamental mode as a function of the string lengths for three different widths. The solid lines correspond to fits using the model used in the main text. \textbf{c} Q of the fundamental mode as a function of the string widths for three different lengths. The gray solid lines correspond to the FEM simulations only for visual comparison. The number written on each line is the vertical to horizontal distributions ratio, $\rho_{e,s}^v/\rho_{e,s}^h$.}
\label{fig:OPT_width_sweep}
\end{figure*}
Throughout the main text, we assumed uniform charge surface densities on all the surfaces of the strings. While this assumption is sufficient to obtain average/order of magnitude values, there is no reason that the charge distribution, at least over the cross section of the devices, is uniform. For example. As illustrated in \subfref{fig:OPT_width_sweep}{a}, the charge distribution on the horizontal and vertical surfaces of the strings ($\rho_{e,s}^h$ and $\rho_{e,s}^v$ respectively) can be different. This can be due to the fact that the vertical surfaces have been subjected to plasma etching (used for patterning the devices) which can introduce surface defects. The charge density difference manifests itself as the dependence of NCF on the width of the strings. \subfref{fig:OPT_width_sweep}{b} shows a measurement of the Q for fundamental modes of strings above the \ce{SiO2} substrate with different lengths, for three families of devices with different widths. The increasing Q for wider strings can be explained by the toy model of section \ref{app:toy_model} and the horizontal/vertical density difference. If the vertical density is much larger than the horizontal one ($\rho_{e,s}^v\gg\rho_{e,s}^h$), wider beams effectively have the same linear charge density but they have a higher mass, resulting in higher Q.

To study in more detail, we fabricate devices with fixed lengths and varying widths in the 2-\SI{7}{\micro\meter} range and measure the Q for their fundamental modes. The results for 4 length families are shown in \subfref{fig:OPT_width_sweep}{c}. As expected, the longer strings have lower Qs, confirming the significant contribution of NCF in these devices. We also observe an overall increasing dependence on the width. Since the widths of these devices becomes bigger than the string-substrate gap (i.e. \SI{500}{\nano\meter}) we cannot use the toy model and use the FEM simulations to study the width dependence. We also show these results in \subfref{fig:OPT_width_sweep}{c}, for different values of $\rho_{e,s}^v/\rho_{e,s}^h$ ratios. We observe that for $\rho_{e,s}^v/\rho_{e,s}^h>2$, the dependence of Q on width becomes monotonically increasing. The experimental data and the FEM results are superimposed in \subfref{fig:OPT_width_sweep}{c} for visual comparison. Although the data quality and the agreement with the model are not very consistent, the qualitative agreement for the trends suggests that the charge density on the vertical sidewalls is higher, which could be explained by the surface damage from plasma etching.

\section{Geometry parameters for the binary-tree resonators}\label{app:nanobeam_design}
\begin{figure}[t!]
\centering
\includegraphics[]{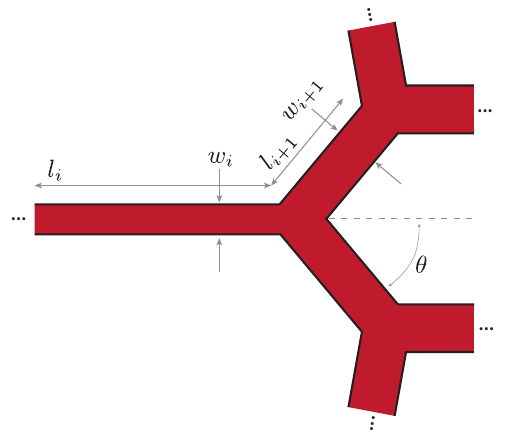}
\caption{\textbf{Geometry parametrization of the binary-tree resonators} The illustration shows a branching point in the resonator. Parameters are described in Table \ref{tab:binary_tree_parameters}. For more details see Ref \cite{fedorov_fractal-like_2020}.}
\label{fig:Binary_tree_details}
\end{figure}
\fref{fig:Binary_tree_details} shows the schematic of a segment of a binary-tree resonator alongside the quantities that parametrize its geometry. These parameters are described in Table \ref{tab:binary_tree_parameters}. These parameters in addition to the \ce{Si3N4} film parameters (thickness and stress) are used to simulate the mechanical mode using COMSOL and compute the properties (Frequency, Q and effective mass) of the first few mechanical modes. We also computes the NCF interaction factor given by \eref{eq:eta_m_int} (See \fref{fig:appendix_HYB_NCF_overlap}). The geometry parameters for different devices used in the experiments are shown in Table \ref{tab:fractal_beams_parameters}

\begin{table*}[h!]
    \begin{tabular}{|c|c|}
        \hline
        Symbol & Description\\
        \hline
        $l_0$ & Length of the central segment\\
        \hline
        $r_l$ & Length expansion ratio $l_{i+1}/l_i$\\
        \hline 
        $w_0$ & Width of the central branch\\
        \hline
        $r_w$ & Width expansion ratio $w_{i+1}/w_i$\\
        \hline
        $\theta$ & Branching angle\\
        \hline
        $N$ & Number of branchin generations\\
        \hline
        $w_p$ & Width of the interaction region\\
        \hline
    \end{tabular}   
    \caption{Geometry parameters of the binary-tree resonators}
    \label{tab:binary_tree_parameters}
\end{table*}

\begin{table*}[t]
    \begin{tabular}{|c|c|c|c|c|c|c|c|c|c|}
        \hline
        Experiment&Run \# &Chip \#& $l_0$ (\SI{ }{\micro\meter}) & $r_l$ & $w_0$ (\SI{ }{\nano\meter})& $r_w$ & $\theta$ & $N$ & $w_p$ (\SI{ }{\nano\meter})\\
        \hline
        Thermal force&1&1 & 250 & 0.59 & 210 & 1.1 & 82 & 2 & 420\\
        \hline
        Gap sweep&1&1 & 250 & 0.59 & 210 & 1.1 & 82 & 2 & 430\\
        \hline
        Mode shape dependence&2&1 & 250 & 0.60 & 210 & 1.27 & 79 & 3 & 430\\
        \hline
        Frequency dependence&3&1 & 250 & 0.6 & 210 & 1.3 & 83 & 2 & 430\\
        \hline
        Frequency dependence&3&2 & 200 & 0.65 & 210 & 1.18 & 79 & 3 & 430\\
        \hline
        Frequency dependence&4&3 & 150 & 0.63 & 210 & 1.2 & 81 & 2 & 430\\
        \hline
    \end{tabular}   
    \caption{Parameters table of different resonators used in the experiments}
    \label{tab:fractal_beams_parameters}
\end{table*}

\section{Fabrication processes}\label{app:fabrication_details}
\begin{figure}[t!]
\includegraphics{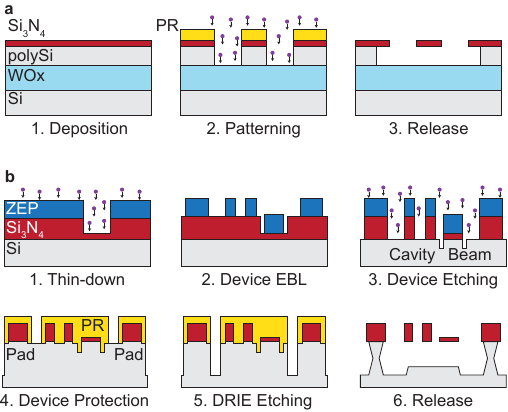}
\caption{\textbf{Fabrication process flows. a} Uniform strings on oxide samples. The initial stack consists of \textit{wet} SiO$_2$ (WOx, 4$\mu$m), polySi (600 nm), and stoichiometric high-stress Si$_3$N$_4$ (200 nm). Nanobeams are patterned in the Si$_3$N$_4$ using optical photolithography and suspended through a combination of vertical and isotropic dry etching. \textbf{b} Integrated nano-optomechanical devices. A 250 nm-thick stoichiometric Si$_3$N$_4$ layer is locally thinned to 20 nm in selected areas using e-beam lithography (EBL). Nanobeams are then patterned in the thinned regions, while photonic crystal (PhC) cavities and supporting pads are defined in the thicker areas, also via EBL. The structures are subsequently protected with photoresist (PR) and suspended using a combination of dry and wet etching.}
\label{fig:fab_pf_opt}
\end{figure}
\subsection{Uniform strings on oxide samples}
An initial layer of \textit{wet} oxide is grown on the Si $<100>$ substrate through wet thermal oxidation, serving both as a dielectric body for the NCF measurements and as an etch-stop layer during fabrication. A sacrificial polySi layer ($\approx$ 600 nm) and a high-stress ($\approx$ 1.1 GPa) stoichiometric Si$_3$N$_4$ device layer ($\approx$ 200 nm) are then deposited via low-pressure chemical vapor deposition (LPCVD). The devices are patterned using optical lithography with a maskless aligner, followed by dry etching of both the device and the sacrificial layers using fluorine chemistry (SF$_6$). This plasma etching process likely induces charging in the devices, leading to the noncontact friction effects described earlier.\newline
The well-controlled deposition rate of polySi LPCVD, combined with the high selectivity ($\approx$ 20:1) of the etching process, ensures precise control of the gap size—a key factor in the findings presented in this paper. Finally, the wafers are diced, and chip-scale release is performed using XeF$_2$ isotropic etching. For these patterns, the XeF$_2$ selectivity for polySi over Si$_3$N$_4$ is approximately 330:1, largely depending on the exposed polySi area. The etching of polySi generates radical byproducts in the form of SiF$_x$, which in turn attack Si$_3$N$_4$ \cite{drysdale_vapour_2015}. Nonetheless, the selectivity remains sufficient to suspend the structures with reasonable reproducibility.\newline
The survival yield of these devices strongly depends on their length and width, reaching unity for beams wider than 2 $\mu$m and increasing as length decreases for narrower beams. The lower yield observed in narrower beams is attributed to their reduced stiffness, making them more susceptible to collapse due to electrical charging effects \cite{cupertino_centimeter-scale_2024}.

\subsection{Integrated nano-optomechanical devices} 
The fabrication process begins on a Si $<100>$ substrate with a 250 nm-thick high-stress stoichiometric Si$_3$N$_4$ layer deposited via low-pressure chemical vapor deposition (LPCVD), sourced from Hahn-Schickard-Gesellschaft. Nanobeam channels are first patterned and locally thinned to 20 nm using e-beam lithography and fluorine-based dry etching. A second e-beam lithography step defines the nanobeams, photonic crystal cavities, and supporting pads, which are then etched using fluorine plasma. Both plasma processes can contribute to electrical charging of the Si$_3$N$_4$ structures.\newline
Following patterning, the devices are protected with PR, and deep reactive ion etching (DRIE) is performed to facilitate release. At this stage, the wafers are diced, and chip-scale release is achieved through potassium hydroxide (KOH) etching, hydrochloric acid (HCl) neutralization, and critical-point drying (CPD). A selectivity on the order of $10^5:1$ is required to successfully release the 20 nm-thick nanobeams.\newline To investigate the possible causes of eventual surface charges on the samples, devices have sometimes been exposed to a 1 minute-long buffered oxide etch right before the CPD step. This is the case, for example, of Chip 3 in \subfref{fig:NCF_OM_analysis}{d}. No reproducible differences in NCF strength were observed when this additional step was performed. \newline
The survival yield primarily depends on the gap size, i.e., the distance between the PhC cavity and the nanobeam. For gaps exceeding 600 nm, the survival yield is above 80\%, whereas for gaps of 200 nm or smaller, it drops below 20\%.

\section{Attempts for mitigation of NCF}\label{App:attempts_for_mitigation}
\subsection{In situ thermal annealing}
It has been experimentally shown that thermal annealing of microfabricated devices can modify the charge density on them \cite{heritier_spatial_2021}. We have done thermal annealing on the nano-optomechanical devices under vacuum. First, the samples are annealed at 80 degrees for about 1 hour under vacuum and the quality factors are remeasured. As shown in \fref{fig:thermal_aneal}, while still far from the reference values, we observe a significant improvement in the Qs. We have performed a second annealing experiment on the same chip at 120 degrees for about 1 hour, upon which we observe reduction of the Qs even below the pre-bake values. We cannot draw a strong conclusion from these experiments and more systematic experiments are required to understand the impact of thermal annealing on the charge densities. 

\begin{figure}[t!]
\centering
\includegraphics[]{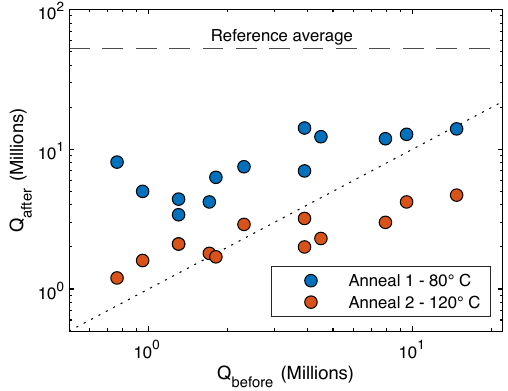}
\caption{\textbf{Modification of the Q upon thermal annealing.} Circles correspond to the measured quality factor after thermal annealing against the pre-bake value. The dashed line shows the average value for the reference (no PhC cavity) devices and the dotted line is an $y=x$ line for aiding the comparison.}
\label{fig:thermal_aneal}
\end{figure}

\subsection{Ultraviolet annealing}
It has also been shown that UV irradiation can modify the charge density in \ce{Si3N4} \cite{sharma_study_2013, krick_nature_1988}. We have tested this method also on a nano-optomechanical device. We have used a \SI{250}{\nano\meter} UV LED (thorlabs LED250J) with above \SI{1}{\milli\watt} power. We illuminated the device at ambient pressure for about 5 minutes and measured the quality factor immediately afterwards. We have observed no significant improvement. The previous studies have shown that much higher power levels are required to make a significant difference \cite{krick_nature_1988}.

\end{widetext}
\end{document}